\documentclass[%
 reprint,superscriptaddress,
 amsmath,amssymb,longbibliography,aps,prl
]{revtex4-2}

\usepackage{natbib}
\usepackage{graphicx}
\usepackage{dcolumn}
\usepackage{bm}
\usepackage{appendix}
\usepackage{graphicx}
\usepackage{dcolumn}
\usepackage{bm}
\usepackage[colorlinks=true,allcolors=blue]{hyperref}
\usepackage{xcolor}
\usepackage{amsmath, amssymb, amsfonts}
\usepackage{mathtools}
\usepackage{subfigure}
\usepackage{mathrsfs}
\usepackage{comment}
\usepackage{amsthm}
\usepackage{ dsfont }

\definecolor{purple1}{rgb}{128,0,128}

\newcommand{\nn}{\nonumber\\}
\newcommand{\bea}{\begin{eqnarray}}
\newcommand{\ea}{\end{eqnarray}}
\newcommand{\ord}{{\cal O}}

\theoremstyle{definition}

\definecolor{dfcol}{cmyk}{1, 0.2108, 0.13, 0.3}
\newcommand{\df}[1]{\ifthenelse{\boolean{}}{\textcolor{dfcol}{[{\bf DF}: #1]}}{}}

\renewcommand{\[}{\begin{equation}}
\renewcommand{\]}{\end{equation}}

\newcommand{\ket}[1]{|#1\rangle}
\newcommand{\bra}[1]{\langle#1|}
\newcommand{\braket}[2]{\langle#1|#2\rangle}

\newcommand{\abs}[1]{\left|#1\right|}

\newcommand{\F}{\mathcal{F}}

\definecolor{mygray}{gray}{0.6}

\begin{document}


\title{Quantum Many-Body Metrology of Rotation Sensing with Strong Interactions}


\author{Yong-Jun Baak}
\author{Uwe R. Fischer}
\affiliation{Seoul National University, Department of Physics and Astronomy, Center for Theoretical Physics, Seoul 08826, Korea}


\date{\today}

\begin{abstract}
We study the ultimate quantum limit of rotation sensing with a few 
strongly interacting bosons confined in a quasi-one-dimensional ring trap with 
two weak links.  
It is demonstrated that a self-consistent many-body solution of the problem is required to correctly predict the ultimate sensitivity of this strongly correlated many-body gyroscope to rotation. 
For both small rotation velocities  
and small  particle numbers, the many-body quantum Fisher information becomes maximal for large interaction couplings, showing the potential of strongly interacting miniaturized many-body sensors to precisely estimate slow rotations with high spatial resolution. 
\end{abstract}


\maketitle


The classic Sagnac experiment, whose sensitivity to rotation scales linearly with the area 
enclosed by the interfering circular light paths and the rotation rate~\cite{SagnacI,SagnacII},  
has with the advent of ultracold quantum gases also been realized with atomic matter waves 
\cite{Pritchard,Kasevich,Gautier,Moan,Gauguet,Dutta}. 
Subsequently, persistent superfluid currents 
have been established \cite{PhysRevLett.106.130401,Ryu,Polo}, 
and SQUID-type cold atom sensors 
based on Josephson-effect weak links \cite{PhysRevLett.110.025302,PhysRevLett.111.205301}, 
similar to rotation sensors realized previously in superfluid helium \cite{Packard,Schwab,Eric}, have been 
implemented. These cold atom sensors 
can be accurately described within mean-field theory~\cite{Zapata,Raghavan,Haine}.  
More recently, fostered by the promise of quantum metrology and sensing to access physical parameters
with unprecedented estimation precision \cite{braunsteinStatisticalDistanceGeometry1994,Braunstein1996,Vittorio2011,Toth_2014,Pirandola},  
 interacting trapped systems of ultracold bosons have been used to establish quantum metrology with 
  highly nonclassical states  whose description is beyond the reach of mean-field theory 
 \cite{Strobel,Gross2010,RevModPhys.90.035005,Klempt,Treutlein,Szigeti}. 
  
\begin{figure}[b]
    \centering
   \vspace*{-2em} 
    \includegraphics[width=\linewidth]{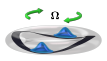}\vspace*{-1em}
    \caption{Illustration of the rotation sensing setup. The  interacting 
    bosons are placed in a strongly confining ring trap,  on which an azimuthal two-site optical lattice 
    potential $V(\theta) = V_0\cos^2\theta $, breaking azimuthal invariance, is superimposed (black), leading to  
localization of the 
single-particle density (blue) for large $V_0$.\label{schema}}
    \label{fig:setup}
\end{figure}

We propose in what follows a novel paradigm for achieving 
the ultimate quantum limit of rotation sensing with ultracold bosons, 
which is neither based on Sagnac 
phase interference nor on the use of highly entangled input states for metrological protocols. 
Whereas the standard paradigm of 
interferometric schemes in quantum optics 
 suggests that interactions are detrimental for the parameter estimation accuracy 
 because they lead to phase diffusion \cite{SinatraCastin,Grond}, we propose a method of Hamiltonian parameter estimation which, conversely, harnesses strong interactions in the ground state. 
While we employ as an additional enhancing factor the proximity to quantum criticality~\cite{Zanardi,frerot,montenegro,mihailescuCriticalQuantumSensing2026}, the core of our 
approach are {\em self-consistently} determined ground state 
many-body correlations which emerge from the combination of trapping and interactions. 
Our sensing scheme is thus based upon the 
impact of the rotation rate on the ground state many-body wave function of 
the gas in the corotating frame and the resulting quantum Fisher information (QFI). 
We specifically stress that we do not employ  
highly entangled input states~\cite{Taylor,luoHeisenberglimitedSagnacInterferometer2017},  
which are susceptible to decoherence. 
In our approach, large entanglement is stably generated by strong coupling at low density in 
a rotating quasi-one-dimensional ring trap with two weak links (see for an illustration Fig.~\ref{schema}). 

We compute the many-body wavefunction to the required level of accuracy 
by a fully self-consistent approach of the multiconfigurational Hartree type, which we demonstrate 
to be necessary to accurately predict the QFI in the strongly interacting trapped system. 
We stress that we do not focus on achieving Heisenberg scaling in particle number~\cite{Ou,QuantumMetrology,Demkowicz}, 
which has been extensively treated, also in the context of rotation sensing~ \cite{Taylor,luoHeisenberglimitedSagnacInterferometer2017,Boixo}. 
Our approach is complementary, in that we explore the beneficial impact of strong interactions,  
using their self-consistent incorporation in a trapped many-body system, and in that our system is 
self-sustained and stable, avoiding the fragility of macroscopic quantum superposition 
(``cat") type to achieve a quantum metrological sensing advantage  \cite{BraunRMP}.  
We demonstrate in fact that the metrological sensitivity of our quantum many-body rotation sensor is for 
large enough coupling the higher the smaller the number of particles is.  
We therefore also show that the QFI, which is indeed a highly  nontrivial functional of many-body correlations,  
the more sensitively captures the impact of the Hamiltonian parameter rotation the larger the correlations encapsulated in the  many-body wave function due to strong interaction coupling are.
The quantum metrological analysis of ultimate quantum many-body sensitivity 
to rotation we perform reveals a number of further peculiar features which we will derive and detail below. 
Among these, particularly noteworthy is that our strongly interacting quantum many-body sensor on a ring can detect small rotations very sensitively: The many-body QFI at vanishing rotation rate remains finite, in distinction
to its counterpart for matter waves described within mean-field theory \cite{Haine}. 



{\it System setup.} 
We consider a system of $N$  repulsively interacting bosons confined in a 
ring trap, cf.~Fig.~\ref{schema}. Our simulations are carried out in the two spatial dimensions perpendicular to the rotation axis~\cite{tsatsosResonancesDynamicalFragmentation2015,beinkeManybodyTunnelingDynamics2015},  
assuming strong confinement in the $z$-direction.   
 In the rotating frame, the $N$-body Hamiltonian is then 
 \begin{multline}
\label{eq:Hamiltonian2D}
    \hat{H} = \sum_{j=1}^{N}\left[-\frac{\hbar^2}{2m}\nabla_j^2+V(\mathbf{r}_j)\right]
    +\sum_{j<l}^N {W}(\mathbf{r}_j-\mathbf{r}_l)-\Omega\hat{L}_z ,
\end{multline} 
where $\mathbf{r}_j=(x_j,y_j)$, $\hat{L}_z$ and $\Omega$ are total angular momentum operator 
and angular velocity to be estimated (assumed to point along $\hat z$), respectively. 
The short-range repulsive interaction between the bosons is modeled by a Gaussian 
${W}(\mathbf{r}_j-\mathbf{r}_l)=\frac{\lambda_\mathrm{2D}} {2\pi\sigma^2}\mathrm{exp}\left(-\frac{(\mathbf{r}_j-\mathbf{r}_l)^2}{2\sigma^2}\right)$ where $\lambda_\mathrm{2D}$ is 2D interaction strength and $\sigma$ interaction range.  
For our numerical calculations, we set $\hbar=m=1$ and put $R=2$, $\omega_\perp=10\sqrt{2}$, and $\sigma=0.25$~\cite{beinkeManybodyTunnelingDynamics2015,doganovTwoTrappedParticles2013a,bhowmikImpactTransverseDirection2020}, where we are choosing an arbitrary length scale $L$ to fix the units of any quantity in powers of $L$. 
For example, considering $\mathrm{^{87}Rb}$ atoms and a microscopic length scale $L=1\,\mu$m, the unit of time is $t_0=1.37\times10^{-3}\ \mathrm{s}$, and $R=2\,\mu\mathrm{m}$, $\Omega_0=1.82\times10^2\ \mathrm{s^{-1}}$.
The external potential is a radially harmonic ring trap with a  superimposed axial double well (two-site optical lattice),   
   $ V(r,\theta)=\frac{1}{2}m\omega_{\perp}^2(r-R)^2+V_0\cos^2\theta$, 
with large radial trapping frequency $\omega_\perp$. 
When the chemical potential $\mu\ll\hbar\omega_{\perp}$, 
the radial degree of freedom is frozen so that the system effectively becomes 
quasi-one-dimensional (quasi-1D). Integrating out the profile (assuming it to be Gaussian) in the radial 
direction of cylindrical coordinates then gives the effective 1D interaction strength, $\lambda_\mathrm{1D}\simeq\lambda_\mathrm{2D}/\sqrt{2\pi(a_{\perp}^2+\sigma^2)}$, 
 where $a_{\perp}=\sqrt{\hbar/(m\omega_{\perp})}$ is the harmonic oscillator length~\cite{PethickSmith2008}.
 The confining ring geometry we envisage can be approximately realized in a flattened torus trap \cite{PhysRevLett.106.130401}. 

The quasi-1D system is 
 characterized by two dimensionless quantities: The (bulk)  dimensionless coupling parameter 
$\gamma=\lambda_\mathrm{1D}/n$, where $n=N/(2\pi R)$ is 1D density,  
and lattice depth $V_0$  in units of recoil energy $E_r=\hbar^2k_L^2/(2m)=\hbar^2/(2mR^2)$, $V_0/E_r$.  
With our chosen parameters, we can achieve $\gamma\sim 69$ (for $\lambda_{\rm 1D}\sim 10$ and $N=2$ we have  $\mu/\hbar\omega_\perp\sim 0.1$, so remain safely within quasi-1D).   
The experimentally realized 
Tonks-Girardeau gas of Ref.~\cite{Paredes} has 
$\gamma=5\ldots200$, so that our parameters allow us to probe the corresponding many-body regime. 
Moreover, we define 
\begin{equation} \Omega_0=\frac\hbar{mR^2} \quad\Longleftrightarrow \quad
\oint ({\bm\Omega} \times {\bm r})\cdot d {\bm s} = 2\pi \Omega R^2 \coloneqq 2\pi\frac\hbar m 
\end{equation} 
as our unit of rotation, which thus corresponds 
to one quantized circulation unit of Sagnac flux for a given $R$. 
 
We analyze a Hamiltonian metrology wherein the information on $\Omega$ is contained in the eigenstates of the Hamiltonian. This is  distinct from the conventional ``two arms"  Sagnac-type 
interferometric protocol, where the prepared states are unitarily propagated in the presence of $\Omega$, and the accumulated phases are measured~\cite{luoHeisenberglimitedSagnacInterferometer2017,dengisVortexNOONStates2026}. Instead, we thus focus on the distinguishability of the full quantum many-body eigenstates of Eq.~(\ref{eq:Hamiltonian2D}) with respect to $\Omega$, quantified by the QFI.  
Given a pure state $\ket{\Psi_\Omega}$, which is parametrically dependent on $\Omega$, the QFI is 
\begin{eqnarray}
    \mathcal{F}_Q(\Omega)=4\left(\braket{\partial_\Omega\Psi_\Omega}{\partial_\Omega\Psi_\Omega}-\abs{\braket{\Psi_\Omega}{\partial_\Omega\Psi_\Omega}}^2\right), \label{FQpure}
\end{eqnarray}
which enters the ultimate limit for $\Omega$ estimation 
 in the Cram\'er-Rao bound of quantum statistics~\cite{helstromMinimumMeansquaredError1967b,braunsteinStatisticalDistanceGeometry1994,facchiClassicalQuantumFisher2010}. The latter reads, for an unbiased estimator,  
\begin{eqnarray}
 (\Delta\Omega)^2 \ge \frac{1}{\nu  \mathcal{F}_Q(\Omega)}, \label{CRbound}
 \end{eqnarray} 
 where $\nu$ is the  number of measurements and $(\Delta\Omega)^2$ is the variance of the 
 rotation rate.

{\it Two-mode approximation without self-consistency.}   
We first consider a conventional two-mode approximation for double wells, which is employing fixed Wannier orbitals ~\cite{milburnQuantumDynamicsAtomic1997a,daltonTwoModeTheory2012}. 
Substituting the two-mode expansion of the field operator $\hat{\Psi}(x)=\hat{b}_1\phi_1(x)+\hat{b}_2\phi_2(x)$, where $\hat{b}_j$ is the boson annihilation operator at sites $j=1,2$ gives the two-site single-band Bose-Hubbard (BH) Hamiltonian
\begin{eqnarray}
    \hat{H}_\mathrm{BH}=-J\cos\theta\left(\hat{b}_1^\dagger\hat{b}_2+\hat{b}_2^\dagger\hat{b}_1\right)
    +\frac{U}{2}\sum_{i=1,2} \hat{b}_i^\dagger\hat{b}_i^\dagger\hat{b}_i \hat{b}_i ,
\end{eqnarray}
where $J$ and $U$ are tunneling rate and on-site interaction, respectively. In this approximation, the rotation effect is solely encoded in the Peierls phase $\theta=\pi\Omega/\Omega_0$, through the modulated tunneling $J_\mathrm{eff}=J\cos\theta$~\cite{milburnQuantumDynamicsAtomic1997a,naldesiEnhancingSensitivityRotations2022,poloPersistentCurrentsUltracold2025}. 
The two-particle ($N=2$) problem can be solved exactly, and thus the QFI 
for estimating $\Omega$ takes the analytical form 
\begin{eqnarray}
\label{eq:tma_qfi}
    \mathcal{F}_Q^\mathrm{BH}(\Omega)=\left(\frac{2\pi}{\Omega_0}\right)^2\frac{\alpha^2\sin^2\left(\frac{\pi\Omega}{\Omega_0}\right)}{\left[4\cos^2\left(\frac{\pi\Omega}{\Omega_0}\right)+\alpha^2\right]^2} ,
\end{eqnarray}
where $\alpha=U/(2J)$ measures the ratio of interaction and tunneling. The BH QFI, 
 $\mathcal{F}_Q^\mathrm{BH}(\Omega)$, reaches its minimum $\mathcal{F}_{Q, \mathrm{min}}^\mathrm{BH}=0$ when $\Omega/\Omega_0$
 is integer or zero, while the maximal QFI is acquired at odd half-integer $\Omega/\Omega_0$, 
\begin{eqnarray}
    \mathcal{F}_{Q, \mathrm{max}}^\mathrm{BH}
    =\mathcal{F}_Q^\mathrm{BH}\left(\Omega=\frac{\Omega_0}{2}\right)
    =\left(\frac{2\pi}{\Omega_0}\right)^2\frac{1}{\alpha^2}.
\end{eqnarray}
Since $U\propto\lambda_\mathrm{1D}$, the maximum QFI decays algebraically as $\lambda_{\mathrm{1D}}^{-2}$.  
This suggests that, according to the BH model, interactions always degrade the QFI. 
We show below that on the self-consistent many-body level this is not the case. 
Rather, the QFI increases with interactions until a maximal value depending on the particle number is reached.

We also note that the maximal BH-QFI is achieved 
when $J_\mathrm{eff}\rightarrow0$ ($U/J_\mathrm{eff}\rightarrow\infty$). When the  effective tunneling is 
suppressed each particle is essentially localized at a site, which would correspond for large $N$ to entering the localized Mott from the delocalized superfluid phase in the archettypical quantum phase transition of the BH model~\cite{jakschColdBosonicAtoms1998a,Greiner}.  
For our small $N$ there is no proper critical behavior, and  {\em critical } 
should be more accurately termed {\em transitional},  as determined by distinct transition features in correlation functions. 
The term quantum phase or criticality will thus be used from now on to refer to 
these  finite-size precursors of quantum critical behavior~\cite{sachdevQuantumPhaseTransitions2011}. 

{\it Numerical many-body approach.} 
We solve the full many-body Schr\"odinger equation in a numerically exact way via the multi-configurational time-dependent Hartree method (MCTDH). For a given number of particles $N$ and orbitals $M$, MCTDH variationally computes both the interdependent orbitals and their Fock space occupation distributions in a self-consistent manner based on the Dirac-Frenkel 
principle~\cite{Meyer,streltsovGeneralVariationalManybody2006, alonMulticonfigurationalTimedependentHartree2008, lodeColloquiumMulticonfigurationalTimedependent2020,Roy,Roy_rotation}. Here, we use its MCTDH-X implementation
\cite{linMCTDHXMulticonfigurationalTimedependent2020}. 
It has been benchmarked for the strongly correlated regime 
with the exactly solvable $N=2$ case in a harmonic trap \cite{Gwak}. 

The many-body QFI can then be computed in a self-consistent manner using the MCTDH-obtained orbitals and Fock space coefficients~\cite{baakSelfConsistentManyBodyMetrology2024}. 
For Josephson junctions (double wells), the exact many-body MCTDH dynamics has been demonstrated 
to deviate strongly from commonly employed 
methods such as 
the BH model, demonstrating its vital importance for predicting the details of many-body states up to large correlation orders \cite{Sakmann,Klaiman}. 
That the details of the many-body state play a crucial role for the QFI 
 is also manifest from the explicit formulas for the QFI \eqref{eq:qfi_total} and \eqref{eq:qfi_orb} we 
 derive in Section A of the End Matter. They
demonstrate a complex interplay between the dependences 
of the orbitals' shape and their occupation statistics. 
The QFI and thus the ultimate parameter estimation accuracy via \eqref{CRbound} are therefore 
strongly affected by self-consistency 
 because the QFI acts as a ``metrological microscope" to enhance  details of the many-body state.


{\it Analysis of the many-body QFI.}
\begin{figure}[t]
    \centering
    \includegraphics[width=\linewidth]{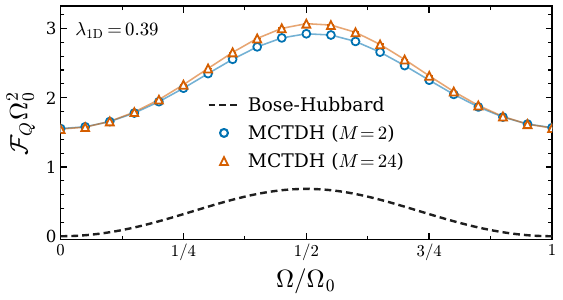}
    \caption{The QFI for $N=2$ bosons with coupling $\lambda_\mathrm{1D}=0.39$  and $V_0/E_r=8$.  
    Dashed line is the BH result with fixed Wannier orbitals. Scattered points show MCTDH calculation results for $M=2$ (circles) and $M=24$ (triangles) orbitals. Lines are guides to the eye. 
    }
    \label{fig:qfi_omega}
\end{figure}
We calculated the QFI 
in the MCTDH framework at an intermediate lattice depth ($V_0/E_r=8$) and compared it with the two-mode approximation in Fig.~\ref{fig:qfi_omega}, for an intermediate  
$\lambda_\mathrm{1D}=0.39$, using $M=2$ and $M=24$ orbitals. 
We first observe that the BH model always underestimates the QFI compared to using $M=2$ variational orbitals.
We verified that also for larger couplings (while staying within the quasi-1D regime), the difference using
$M=2$ or $M=24$ orbitals (for $N=2$) remains in the percent range. 

A primary result of our calculations is that MCTDH 
predicts a nonzero and large value of the QFI at $\Omega=0$, while the two-mode BH model gives exactly zero. 
This strikingly demonstrates the vital importance of self-consistency for metrological accuracy in our setup. 
The QFI, also for $M>2$, generally reaches its maximum at $\Omega=\Omega_0/2$ and minimum at $\Omega=0$, 
a pattern recurring with period $\Omega_0$. 
\begin{figure}[b]
    \centering
    \includegraphics[width=\linewidth]{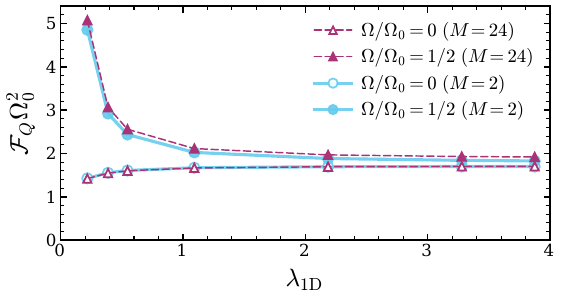}
    \caption{Convergence of QFI for large coupling, 
    for $\Omega=0$ (from below) and $\Omega_0/2$ (from above), with $M=2$ and $24$ self-consistently 
    computed orbitals ($N=2$).} 
    \label{fig:qfi_gamma}
\end{figure}
The QFI at $\Omega=0$ and $\Omega_0/2$ is displayed for varying interaction couplings $\lambda_{\rm 1D}$  in Fig.~\ref{fig:qfi_gamma}. As the interaction strength is increased, $\mathcal{F}_Q(\Omega=0)$ becomes larger and saturates at a finite value. On the other hand, the maximum QFI at $\Omega=\Omega_0/2$ decreases, converging to a saturated value higher than at $\Omega=0$. 
As a corollary, a QFI calculated with 
  {\em self-consistently} determined, interaction-broadened Wannier-type orbitals also would give 
   within a correspondingly amended single-band ($M=2$) BH model a fair estimate, at least in the range of interaction couplings investigated.  

\begin{figure}[t]
    \centering
    \includegraphics[width=\linewidth]
    {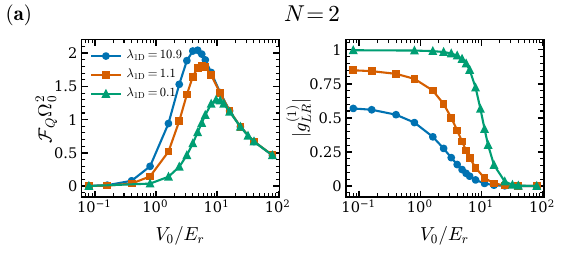}
    \includegraphics[width=\linewidth]{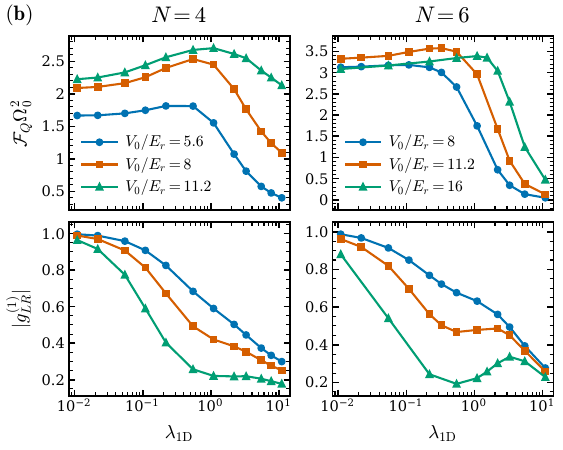}
    \caption{(a) Left panel: Lattice-depth dependence of the QFI at $\Omega=0$ for $N=2$ interacting bosons. Right panel:  Normalized one-body Glauber correlation function $g^{(1)}_{LR}=g^{(1)}(\theta=\pi/2, \theta'=3\pi/2)$ between the two azimuthal lattice sites. $M=8$ orbitals were used for the self-consistent calculation. (b) Interaction-strength dependence of the QFI at 
    $\Omega=0$ for $N=4$ (left column) and $N=6$ (right column) bosons. 
    $M=12$ ($M=18$) orbitals were used for $N=4$ ($N=6$) to reach convergence.}  
    \label{fig:qfi_corr_N_2}
\end{figure}

{\it Criticality-enhanced QFI.} 
We now investigate the QFI in the self-consistent framework by varying $V_0/E_r$ to demonstrate criticality-enhancement of the QFI.  Fig.~\ref{fig:qfi_corr_N_2}(a) left panel shows the QFI at $\Omega=0$ for $N=2$ bosons and for various lattice depths. As a measure of the first-order many-body coherence, we consider the normalized one-body azimuthal Glauber (first-order) correlation function~\cite{glauberQuantumTheoryOptical1963, royPhasesManybodyEntropy2018,linSuperfluidMottInsulator2019} defined as follows 
\begin{equation}
g^{(1)} (\theta,\theta') 
=\frac{\bra{\Psi}\hat{\Psi}(\theta)\hat{\Psi}(\theta')\ket{\Psi}}{N\sqrt{\rho(\theta)\rho(\theta')}}.
\end{equation} 
The corresponding spatial correlation is computed between the two lattice sites 
$g^{(1)}_{LR}\coloneqq 
g^{(1)}(\theta=\pi/2, \theta'=3\pi/2)$ and is shown for $N=2$ 
in Fig.~\ref{fig:qfi_corr_N_2}(a) on the right panel. 
We have verified that the second-order correlations $g^{(2)}$ show
a similar transitional behavior except they increase rather then decrease towards the localized phase.  

For a given interaction strength, the QFI displays a distinct peak at some intermediate lattice depth $(V_0/E_r)_{\mathrm{max}}$, near which the correlation $|g^{(1)}_{LR}|$ decreases steeply. For lattice depths lower than $(V_0/E_r)_{\mathrm{max}}$, $|g^{(1)}_{LR}|$ maintains a non-zero value, indicating a phase correlation between the two lattice sites, and a superfluid (delocalized) phase~\cite{boerisMottTransitionStrongly2016,royPhasesManybodyEntropy2018, linSuperfluidMottInsulator2019}. As the lattice depth is further increased, the off-diagonal correlation decreases steeply near $(V_0/E_r)_{\mathrm{max}}$, and ultimately vanishes ($|g^{(1)}_{LR}|\simeq0$), entering a Mott-insulating localized phase. Therefore, the maximum of the QFI at $\Omega=0$ is acquired in the vicinity of the superfluid-to-Mott-insulator transition. Maximization of the QFI near quantum criticality has been widely reported in other systems as well~\cite{Zanardi,frerot,montenegro,mihailescuCriticalQuantumSensing2026}. 

To demonstrate the impact of coupling on criticality-enhanced QFI, we also computed the latter for $N=4$ and $N=6$ at vanishing rotation rate,  $\Omega=0$, and for a range of $\lambda_\mathrm{1D}$ values and three different values 
of $V_0/E_r$.   Fig.~\ref{fig:qfi_corr_N_2}(b) shows the QFI (first row) together with the normalized one-body Glauber correlation function $|g^{(1)}_{LR}|$ (second row). 
When $\gamma$ is increased, the QFI peak height increases and is shifted towards shallower lattice depths. 
The decrement of the critical well depth is expected for repulsively interacting bosons trapped in a two-site optical lattice  (corresponding to two weak links), as the localization can occur even at a slight increase of well depth when the interaction is strong enough~\cite{boerisMottTransitionStrongly2016}. 
We note here that 
stronger interaction always helps sensitivity at $\Omega=0$ for $N=2$ particles, cf.~Fig.~\ref{fig:qfi_corr_N_2}(a) left panel. 
Operating within our confines of the quasi-1D limit 
(which as already mentioned still allows access to the Tonks-Girardeau gas \cite{Paredes}),  
for both $N=4$ and $N=6$, the maximal QFI is obtained at an intermediate interaction strength (close to the ``fermionization"  crossover), which is larger for deeper lattices $V_0/E_r$ [Fig.~\ref{fig:qfi_corr_N_2}(b)]. 
This is consistent with the fact that 
stronger interactions are required for the bosons to ``fermionize" in deeper potential  wells~\cite{alonPathwayCondensationFragmentation2005}. 
This transition behavior of the QFI 
also appears when the parameter to be estimated is the coupling constant itself \cite{LiebLinigerMetro}.  
Finally, to demonstrate the effects of increasing the coupling 
when more than one boson occupies a well, we discuss in the End Matter, Section B, 
the crossover to a ``fermionized" Tonks-Girardeau gas in each well \cite{TG,Paredes} for $N>2$ and $N$ even (commensurate filling).

{\it Conclusion.}  
We have conducted a self-consistent many-body analysis of the quantum metrology of 
rotation sensing as a Hamiltonian parameter estimation problem 
with an ultracold strongly interacting quasi-1D Bose gas confined in a ring trap. 
Specifically, we have found that increasing the interaction coupling 
enhances for small rotation rates  
the ultimate quantum sensitivity encapsulated in the QFI at vanishing $\Omega$,  
$\mathcal{\F}_Q(\Omega\rightarrow 0)=\ord{\Omega_0^2}$, which remains finite due to many-body effects. 
Our results therefore demonstrate the increased sensitivity of strongly correlated quantum many-body states to rotation over their mean-field counterparts, as encoded in the magnifying glass of ${\mathcal F}_Q(\Omega)$. 

Modern quantum gas microscope techniques 
 \cite{Bakr,Gross,Zwierlein}  enable the determination of the 
particle distribution statistics 
even for very  small particle numbers. 
 The classical Fisher information corresponding to measuring such particle distribution functions in a double well  
 has been demonstrated to be close to the QFI, when tilting a double well for gravity estimation 
 \cite{baakSelfConsistentManyBodyMetrology2024}, and we expect a similar result for rotation sensing.

We envisage applications of our results in the emerging field of atomtronics \cite{Amico},  
more specifically for constructing a spatially highly resolved sensing of inhomogeneous rotation fields.  
One may to this end arrange many tiny ring traps, each hosting a few strongly interacting atoms, 
with different orientations in a suitably arranged miniaturized design,  to achieve 
a map of inhomogeneous rotation fields with high spatial resolution, akin to what has been achieved for imaging brain activity in magnetoencephalography with SQUIDs~\cite{Lounasmaa}. 

This work was still supported by the 
NRF of Korea under 
Grant No.~2020R1A2C2008103. It has not been supported by the 
IRC NextQuantum at Seoul National University. 

\bibliography{qmbrs7}

\begin{thebibliography}{84}%
\makeatletter
\providecommand \@ifxundefined [1]{%
 \@ifx{#1\undefined}
}%
\providecommand \@ifnum [1]{%
 \ifnum #1\expandafter \@firstoftwo
 \else \expandafter \@secondoftwo
 \fi
}%
\providecommand \@ifx [1]{%
 \ifx #1\expandafter \@firstoftwo
 \else \expandafter \@secondoftwo
 \fi
}%
\providecommand \natexlab [1]{#1}%
\providecommand \enquote  [1]{``#1''}%
\providecommand \bibnamefont  [1]{#1}%
\providecommand \bibfnamefont [1]{#1}%
\providecommand \citenamefont [1]{#1}%
\providecommand \href@noop [0]{\@secondoftwo}%
\providecommand \href [0]{\begingroup \@sanitize@url \@href}%
\providecommand \@href[1]{\@@startlink{#1}\@@href}%
\providecommand \@@href[1]{\endgroup#1\@@endlink}%
\providecommand \@sanitize@url [0]{\catcode `\\12\catcode `\$12\catcode
  `\&12\catcode `\#12\catcode `\^12\catcode `\_12\catcode `\%12\relax}%
\providecommand \@@startlink[1]{}%
\providecommand \@@endlink[0]{}%
\providecommand \url  [0]{\begingroup\@sanitize@url \@url }%
\providecommand \@url [1]{\endgroup\@href {#1}{\urlprefix }}%
\providecommand \urlprefix  [0]{URL }%
\providecommand \Eprint [0]{\href }%
\providecommand \doibase [0]{https://doi.org/}%
\providecommand \selectlanguage [0]{\@gobble}%
\providecommand \bibinfo  [0]{\@secondoftwo}%
\providecommand \bibfield  [0]{\@secondoftwo}%
\providecommand \translation [1]{[#1]}%
\providecommand \BibitemOpen [0]{}%
\providecommand \bibitemStop [0]{}%
\providecommand \bibitemNoStop [0]{.\EOS\space}%
\providecommand \EOS [0]{\spacefactor3000\relax}%
\providecommand \BibitemShut  [1]{\csname bibitem#1\endcsname}%
\let\auto@bib@innerbib\@empty
\bibitem [{\citenamefont {Sagnac}(1913{\natexlab{a}})}]{SagnacI}%
  \BibitemOpen
  \bibfield  {author} {\bibinfo {author} {\bibfnamefont {G.}~\bibnamefont
  {Sagnac}},\ }\bibfield  {title} {\bibinfo {title} {L'éther lumineux
  démontré par l'effet du vent relatif d'éther dans un interféromètre en
  rotation uniforme},\ }\href@noop {} {\bibfield  {journal} {\bibinfo
  {journal} {C. R. Acad. Sci. Paris}\ }\textbf {\bibinfo {volume} {157}},\
  \bibinfo {pages} {708–710} (\bibinfo {year}
  {1913}{\natexlab{a}})}\BibitemShut {NoStop}%
\bibitem [{\citenamefont {Sagnac}(1913{\natexlab{b}})}]{SagnacII}%
  \BibitemOpen
  \bibfield  {author} {\bibinfo {author} {\bibfnamefont {G.}~\bibnamefont
  {Sagnac}},\ }\bibfield  {title} {\bibinfo {title} {Sur la preuve de la
  réalité de l'éther lumineux par l'expérience de l'interférographe
  tournant},\ }\href@noop {} {\bibfield  {journal} {\bibinfo  {journal} {C. R.
  Acad. Sci. Paris}\ }\textbf {\bibinfo {volume} {157}},\ \bibinfo {pages}
  {1410–1413} (\bibinfo {year} {1913}{\natexlab{b}})}\BibitemShut {NoStop}%
\bibitem [{\citenamefont {Lenef}\ \emph {et~al.}(1997)\citenamefont {Lenef},
  \citenamefont {Hammond}, \citenamefont {Smith}, \citenamefont {Chapman},
  \citenamefont {Rubenstein},\ and\ \citenamefont {Pritchard}}]{Pritchard}%
  \BibitemOpen
  \bibfield  {author} {\bibinfo {author} {\bibfnamefont {A.}~\bibnamefont
  {Lenef}}, \bibinfo {author} {\bibfnamefont {T.~D.}\ \bibnamefont {Hammond}},
  \bibinfo {author} {\bibfnamefont {E.~T.}\ \bibnamefont {Smith}}, \bibinfo
  {author} {\bibfnamefont {M.~S.}\ \bibnamefont {Chapman}}, \bibinfo {author}
  {\bibfnamefont {R.~A.}\ \bibnamefont {Rubenstein}},\ and\ \bibinfo {author}
  {\bibfnamefont {D.~E.}\ \bibnamefont {Pritchard}},\ }\bibfield  {title}
  {\bibinfo {title} {{Rotation Sensing with an Atom Interferometer}},\ }\href
  {https://doi.org/10.1103/PhysRevLett.78.760} {\bibfield  {journal} {\bibinfo
  {journal} {Phys. Rev. Lett.}\ }\textbf {\bibinfo {volume} {78}},\ \bibinfo
  {pages} {760} (\bibinfo {year} {1997})}\BibitemShut {NoStop}%
\bibitem [{\citenamefont {Gustavson}\ \emph {et~al.}(1997)\citenamefont
  {Gustavson}, \citenamefont {Bouyer},\ and\ \citenamefont
  {Kasevich}}]{Kasevich}%
  \BibitemOpen
  \bibfield  {author} {\bibinfo {author} {\bibfnamefont {T.~L.}\ \bibnamefont
  {Gustavson}}, \bibinfo {author} {\bibfnamefont {P.}~\bibnamefont {Bouyer}},\
  and\ \bibinfo {author} {\bibfnamefont {M.~A.}\ \bibnamefont {Kasevich}},\
  }\bibfield  {title} {\bibinfo {title} {{Precision Rotation Measurements with
  an Atom Interferometer Gyroscope}},\ }\href
  {https://doi.org/10.1103/PhysRevLett.78.2046} {\bibfield  {journal} {\bibinfo
   {journal} {Phys. Rev. Lett.}\ }\textbf {\bibinfo {volume} {78}},\ \bibinfo
  {pages} {2046} (\bibinfo {year} {1997})}\BibitemShut {NoStop}%
\bibitem [{\citenamefont {Gautier}\ \emph {et~al.}(2022)\citenamefont
  {Gautier}, \citenamefont {Guessoum}, \citenamefont {Sidorenkov},
  \citenamefont {Bouton}, \citenamefont {Landragin},\ and\ \citenamefont
  {Geiger}}]{Gautier}%
  \BibitemOpen
  \bibfield  {author} {\bibinfo {author} {\bibfnamefont {R.}~\bibnamefont
  {Gautier}}, \bibinfo {author} {\bibfnamefont {M.}~\bibnamefont {Guessoum}},
  \bibinfo {author} {\bibfnamefont {L.~A.}\ \bibnamefont {Sidorenkov}},
  \bibinfo {author} {\bibfnamefont {Q.}~\bibnamefont {Bouton}}, \bibinfo
  {author} {\bibfnamefont {A.}~\bibnamefont {Landragin}},\ and\ \bibinfo
  {author} {\bibfnamefont {R.}~\bibnamefont {Geiger}},\ }\bibfield  {title}
  {\bibinfo {title} {{Accurate measurement of the Sagnac effect for matter
  waves}},\ }\href {https://doi.org/10.1126/sciadv.abn8009} {\bibfield
  {journal} {\bibinfo  {journal} {Science Advances}\ }\textbf {\bibinfo
  {volume} {8}},\ \bibinfo {pages} {eabn8009} (\bibinfo {year}
  {2022})}\BibitemShut {NoStop}%
\bibitem [{\citenamefont {Moan}\ \emph {et~al.}(2020)\citenamefont {Moan},
  \citenamefont {Horne}, \citenamefont {Arpornthip}, \citenamefont {Luo},
  \citenamefont {Fallon}, \citenamefont {Berl},\ and\ \citenamefont
  {Sackett}}]{Moan}%
  \BibitemOpen
  \bibfield  {author} {\bibinfo {author} {\bibfnamefont {E.~R.}\ \bibnamefont
  {Moan}}, \bibinfo {author} {\bibfnamefont {R.~A.}\ \bibnamefont {Horne}},
  \bibinfo {author} {\bibfnamefont {T.}~\bibnamefont {Arpornthip}}, \bibinfo
  {author} {\bibfnamefont {Z.}~\bibnamefont {Luo}}, \bibinfo {author}
  {\bibfnamefont {A.~J.}\ \bibnamefont {Fallon}}, \bibinfo {author}
  {\bibfnamefont {S.~J.}\ \bibnamefont {Berl}},\ and\ \bibinfo {author}
  {\bibfnamefont {C.~A.}\ \bibnamefont {Sackett}},\ }\bibfield  {title}
  {\bibinfo {title} {{Quantum Rotation Sensing with Dual Sagnac Interferometers
  in an Atom-Optical Waveguide}},\ }\href
  {https://doi.org/10.1103/PhysRevLett.124.120403} {\bibfield  {journal}
  {\bibinfo  {journal} {Phys. Rev. Lett.}\ }\textbf {\bibinfo {volume} {124}},\
  \bibinfo {pages} {120403} (\bibinfo {year} {2020})}\BibitemShut {NoStop}%
\bibitem [{\citenamefont {Gauguet}\ \emph {et~al.}(2009)\citenamefont
  {Gauguet}, \citenamefont {Canuel}, \citenamefont {L\'ev\`eque}, \citenamefont
  {Chaibi},\ and\ \citenamefont {Landragin}}]{Gauguet}%
  \BibitemOpen
  \bibfield  {author} {\bibinfo {author} {\bibfnamefont {A.}~\bibnamefont
  {Gauguet}}, \bibinfo {author} {\bibfnamefont {B.}~\bibnamefont {Canuel}},
  \bibinfo {author} {\bibfnamefont {T.}~\bibnamefont {L\'ev\`eque}}, \bibinfo
  {author} {\bibfnamefont {W.}~\bibnamefont {Chaibi}},\ and\ \bibinfo {author}
  {\bibfnamefont {A.}~\bibnamefont {Landragin}},\ }\bibfield  {title} {\bibinfo
  {title} {{Characterization and limits of a cold-atom Sagnac
  interferometer}},\ }\href {https://doi.org/10.1103/PhysRevA.80.063604}
  {\bibfield  {journal} {\bibinfo  {journal} {Phys. Rev. A}\ }\textbf {\bibinfo
  {volume} {80}},\ \bibinfo {pages} {063604} (\bibinfo {year}
  {2009})}\BibitemShut {NoStop}%
\bibitem [{\citenamefont {Dutta}\ \emph {et~al.}(2016)\citenamefont {Dutta},
  \citenamefont {Savoie}, \citenamefont {Fang}, \citenamefont {Venon},
  \citenamefont {Garrido~Alzar}, \citenamefont {Geiger},\ and\ \citenamefont
  {Landragin}}]{Dutta}%
  \BibitemOpen
  \bibfield  {author} {\bibinfo {author} {\bibfnamefont {I.}~\bibnamefont
  {Dutta}}, \bibinfo {author} {\bibfnamefont {D.}~\bibnamefont {Savoie}},
  \bibinfo {author} {\bibfnamefont {B.}~\bibnamefont {Fang}}, \bibinfo {author}
  {\bibfnamefont {B.}~\bibnamefont {Venon}}, \bibinfo {author} {\bibfnamefont
  {C.~L.}\ \bibnamefont {Garrido~Alzar}}, \bibinfo {author} {\bibfnamefont
  {R.}~\bibnamefont {Geiger}},\ and\ \bibinfo {author} {\bibfnamefont
  {A.}~\bibnamefont {Landragin}},\ }\bibfield  {title} {\bibinfo {title}
  {{Continuous Cold-Atom Inertial Sensor with $1\text{ }\text{
  }\mathrm{nrad}/\mathrm{sec}$ Rotation Stability}},\ }\href
  {https://doi.org/10.1103/PhysRevLett.116.183003} {\bibfield  {journal}
  {\bibinfo  {journal} {Phys. Rev. Lett.}\ }\textbf {\bibinfo {volume} {116}},\
  \bibinfo {pages} {183003} (\bibinfo {year} {2016})}\BibitemShut {NoStop}%
\bibitem [{\citenamefont {Ramanathan}\ \emph {et~al.}(2011)\citenamefont
  {Ramanathan}, \citenamefont {Wright}, \citenamefont {Muniz}, \citenamefont
  {Zelan}, \citenamefont {Hill}, \citenamefont {Lobb}, \citenamefont
  {Helmerson}, \citenamefont {Phillips},\ and\ \citenamefont
  {Campbell}}]{PhysRevLett.106.130401}%
  \BibitemOpen
  \bibfield  {author} {\bibinfo {author} {\bibfnamefont {A.}~\bibnamefont
  {Ramanathan}}, \bibinfo {author} {\bibfnamefont {K.~C.}\ \bibnamefont
  {Wright}}, \bibinfo {author} {\bibfnamefont {S.~R.}\ \bibnamefont {Muniz}},
  \bibinfo {author} {\bibfnamefont {M.}~\bibnamefont {Zelan}}, \bibinfo
  {author} {\bibfnamefont {W.~T.}\ \bibnamefont {Hill}}, \bibinfo {author}
  {\bibfnamefont {C.~J.}\ \bibnamefont {Lobb}}, \bibinfo {author}
  {\bibfnamefont {K.}~\bibnamefont {Helmerson}}, \bibinfo {author}
  {\bibfnamefont {W.~D.}\ \bibnamefont {Phillips}},\ and\ \bibinfo {author}
  {\bibfnamefont {G.~K.}\ \bibnamefont {Campbell}},\ }\bibfield  {title}
  {\bibinfo {title} {{Superflow in a Toroidal Bose-Einstein Condensate: An Atom
  Circuit with a Tunable Weak Link}},\ }\href
  {https://doi.org/10.1103/PhysRevLett.106.130401} {\bibfield  {journal}
  {\bibinfo  {journal} {Phys. Rev. Lett.}\ }\textbf {\bibinfo {volume} {106}},\
  \bibinfo {pages} {130401} (\bibinfo {year} {2011})}\BibitemShut {NoStop}%
\bibitem [{\citenamefont {Ryu}\ \emph {et~al.}(2020)\citenamefont {Ryu},
  \citenamefont {Samson},\ and\ \citenamefont {Boshier}}]{Ryu}%
  \BibitemOpen
  \bibfield  {author} {\bibinfo {author} {\bibfnamefont {C.}~\bibnamefont
  {Ryu}}, \bibinfo {author} {\bibfnamefont {E.~C.}\ \bibnamefont {Samson}},\
  and\ \bibinfo {author} {\bibfnamefont {M.~G.}\ \bibnamefont {Boshier}},\
  }\bibfield  {title} {\bibinfo {title} {{Quantum interference of currents in
  an atomtronic SQUID}},\ }\href {https://doi.org/10.1038/s41467-020-17185-6}
  {\bibfield  {journal} {\bibinfo  {journal} {Nature Communications}\ }\textbf
  {\bibinfo {volume} {11}},\ \bibinfo {pages} {3338} (\bibinfo {year}
  {2020})}\BibitemShut {NoStop}%
\bibitem [{\citenamefont {Polo}\ \emph
  {et~al.}(2025{\natexlab{a}})\citenamefont {Polo}, \citenamefont {Chetcuti},
  \citenamefont {Haug}, \citenamefont {Minguzzi}, \citenamefont {Wright},\ and\
  \citenamefont {Amico}}]{Polo}%
  \BibitemOpen
  \bibfield  {author} {\bibinfo {author} {\bibfnamefont {J.}~\bibnamefont
  {Polo}}, \bibinfo {author} {\bibfnamefont {W.}~\bibnamefont {Chetcuti}},
  \bibinfo {author} {\bibfnamefont {T.}~\bibnamefont {Haug}}, \bibinfo {author}
  {\bibfnamefont {A.}~\bibnamefont {Minguzzi}}, \bibinfo {author}
  {\bibfnamefont {K.}~\bibnamefont {Wright}},\ and\ \bibinfo {author}
  {\bibfnamefont {L.}~\bibnamefont {Amico}},\ }\bibfield  {title} {\bibinfo
  {title} {Persistent currents in ultracold gases},\ }\href
  {https://doi.org/https://doi.org/10.1016/j.physrep.2025.06.003} {\bibfield
  {journal} {\bibinfo  {journal} {Physics Reports}\ }\textbf {\bibinfo {volume}
  {1137}},\ \bibinfo {pages} {1} (\bibinfo {year}
  {2025}{\natexlab{a}})}\BibitemShut {NoStop}%
\bibitem [{\citenamefont {Wright}\ \emph {et~al.}(2013)\citenamefont {Wright},
  \citenamefont {Blakestad}, \citenamefont {Lobb}, \citenamefont {Phillips},\
  and\ \citenamefont {Campbell}}]{PhysRevLett.110.025302}%
  \BibitemOpen
  \bibfield  {author} {\bibinfo {author} {\bibfnamefont {K.~C.}\ \bibnamefont
  {Wright}}, \bibinfo {author} {\bibfnamefont {R.~B.}\ \bibnamefont
  {Blakestad}}, \bibinfo {author} {\bibfnamefont {C.~J.}\ \bibnamefont {Lobb}},
  \bibinfo {author} {\bibfnamefont {W.~D.}\ \bibnamefont {Phillips}},\ and\
  \bibinfo {author} {\bibfnamefont {G.~K.}\ \bibnamefont {Campbell}},\
  }\bibfield  {title} {\bibinfo {title} {{Driving Phase Slips in a Superfluid
  Atom Circuit with a Rotating Weak Link}},\ }\href
  {https://doi.org/10.1103/PhysRevLett.110.025302} {\bibfield  {journal}
  {\bibinfo  {journal} {Phys. Rev. Lett.}\ }\textbf {\bibinfo {volume} {110}},\
  \bibinfo {pages} {025302} (\bibinfo {year} {2013})}\BibitemShut {NoStop}%
\bibitem [{\citenamefont {Ryu}\ \emph {et~al.}(2013)\citenamefont {Ryu},
  \citenamefont {Blackburn}, \citenamefont {Blinova},\ and\ \citenamefont
  {Boshier}}]{PhysRevLett.111.205301}%
  \BibitemOpen
  \bibfield  {author} {\bibinfo {author} {\bibfnamefont {C.}~\bibnamefont
  {Ryu}}, \bibinfo {author} {\bibfnamefont {P.~W.}\ \bibnamefont {Blackburn}},
  \bibinfo {author} {\bibfnamefont {A.~A.}\ \bibnamefont {Blinova}},\ and\
  \bibinfo {author} {\bibfnamefont {M.~G.}\ \bibnamefont {Boshier}},\
  }\bibfield  {title} {\bibinfo {title} {{Experimental Realization of Josephson
  Junctions for an Atom SQUID}},\ }\href
  {https://doi.org/10.1103/PhysRevLett.111.205301} {\bibfield  {journal}
  {\bibinfo  {journal} {Phys. Rev. Lett.}\ }\textbf {\bibinfo {volume} {111}},\
  \bibinfo {pages} {205301} (\bibinfo {year} {2013})}\BibitemShut {NoStop}%
\bibitem [{\citenamefont {Packard}\ and\ \citenamefont
  {Vitale}(1992)}]{Packard}%
  \BibitemOpen
  \bibfield  {author} {\bibinfo {author} {\bibfnamefont {R.~E.}\ \bibnamefont
  {Packard}}\ and\ \bibinfo {author} {\bibfnamefont {S.}~\bibnamefont
  {Vitale}},\ }\bibfield  {title} {\bibinfo {title} {Principles of
  superfluid-helium gyroscopes},\ }\href
  {https://doi.org/10.1103/PhysRevB.46.3540} {\bibfield  {journal} {\bibinfo
  {journal} {Phys. Rev. B}\ }\textbf {\bibinfo {volume} {46}},\ \bibinfo
  {pages} {3540} (\bibinfo {year} {1992})}\BibitemShut {NoStop}%
\bibitem [{\citenamefont {Schwab}\ \emph {et~al.}(1997)\citenamefont {Schwab},
  \citenamefont {Bruckner},\ and\ \citenamefont {Packard}}]{Schwab}%
  \BibitemOpen
  \bibfield  {author} {\bibinfo {author} {\bibfnamefont {K.}~\bibnamefont
  {Schwab}}, \bibinfo {author} {\bibfnamefont {N.}~\bibnamefont {Bruckner}},\
  and\ \bibinfo {author} {\bibfnamefont {R.~E.}\ \bibnamefont {Packard}},\
  }\bibfield  {title} {\bibinfo {title} {{Detection of the Earth's rotation
  using superfluid phase coherence}},\ }\href
  {https://doi.org/10.1038/386585a0} {\bibfield  {journal} {\bibinfo  {journal}
  {Nature}\ }\textbf {\bibinfo {volume} {386}},\ \bibinfo {pages} {585}
  (\bibinfo {year} {1997})}\BibitemShut {NoStop}%
\bibitem [{\citenamefont {Avenel}\ \emph {et~al.}(1997)\citenamefont {Avenel},
  \citenamefont {Hakonen},\ and\ \citenamefont {Varoquaux}}]{Eric}%
  \BibitemOpen
  \bibfield  {author} {\bibinfo {author} {\bibfnamefont {O.}~\bibnamefont
  {Avenel}}, \bibinfo {author} {\bibfnamefont {P.}~\bibnamefont {Hakonen}},\
  and\ \bibinfo {author} {\bibfnamefont {E.}~\bibnamefont {Varoquaux}},\
  }\bibfield  {title} {\bibinfo {title} {{Detection of the Rotation of the
  Earth with a Superfluid Gyrometer}},\ }\href
  {https://doi.org/10.1103/PhysRevLett.78.3602} {\bibfield  {journal} {\bibinfo
   {journal} {Phys. Rev. Lett.}\ }\textbf {\bibinfo {volume} {78}},\ \bibinfo
  {pages} {3602} (\bibinfo {year} {1997})}\BibitemShut {NoStop}%
\bibitem [{\citenamefont {Zapata}\ \emph {et~al.}(1998)\citenamefont {Zapata},
  \citenamefont {Sols},\ and\ \citenamefont {Leggett}}]{Zapata}%
  \BibitemOpen
  \bibfield  {author} {\bibinfo {author} {\bibfnamefont {I.}~\bibnamefont
  {Zapata}}, \bibinfo {author} {\bibfnamefont {F.}~\bibnamefont {Sols}},\ and\
  \bibinfo {author} {\bibfnamefont {A.~J.}\ \bibnamefont {Leggett}},\
  }\bibfield  {title} {\bibinfo {title} {{Josephson effect between trapped
  Bose-Einstein condensates}},\ }\href
  {https://doi.org/10.1103/PhysRevA.57.R28} {\bibfield  {journal} {\bibinfo
  {journal} {Phys. Rev. A}\ }\textbf {\bibinfo {volume} {57}},\ \bibinfo
  {pages} {R28(R)} (\bibinfo {year} {1998})}\BibitemShut {NoStop}%
\bibitem [{\citenamefont {Raghavan}\ \emph {et~al.}(1999)\citenamefont
  {Raghavan}, \citenamefont {Smerzi}, \citenamefont {Fantoni},\ and\
  \citenamefont {Shenoy}}]{Raghavan}%
  \BibitemOpen
  \bibfield  {author} {\bibinfo {author} {\bibfnamefont {S.}~\bibnamefont
  {Raghavan}}, \bibinfo {author} {\bibfnamefont {A.}~\bibnamefont {Smerzi}},
  \bibinfo {author} {\bibfnamefont {S.}~\bibnamefont {Fantoni}},\ and\ \bibinfo
  {author} {\bibfnamefont {S.~R.}\ \bibnamefont {Shenoy}},\ }\bibfield  {title}
  {\bibinfo {title} {{Coherent oscillations between two weakly coupled
  Bose-Einstein condensates: Josephson effects, $\ensuremath{\pi}$
  oscillations, and macroscopic quantum self-trapping}},\ }\href
  {https://doi.org/10.1103/PhysRevA.59.620} {\bibfield  {journal} {\bibinfo
  {journal} {Phys. Rev. A}\ }\textbf {\bibinfo {volume} {59}},\ \bibinfo
  {pages} {620} (\bibinfo {year} {1999})}\BibitemShut {NoStop}%
\bibitem [{\citenamefont {Haine}(2016)}]{Haine}%
  \BibitemOpen
  \bibfield  {author} {\bibinfo {author} {\bibfnamefont {S.~A.}\ \bibnamefont
  {Haine}},\ }\bibfield  {title} {\bibinfo {title} {{Mean-Field Dynamics and
  Fisher Information in Matter Wave Interferometry}},\ }\href
  {https://doi.org/10.1103/PhysRevLett.116.230404} {\bibfield  {journal}
  {\bibinfo  {journal} {Phys. Rev. Lett.}\ }\textbf {\bibinfo {volume} {116}},\
  \bibinfo {pages} {230404} (\bibinfo {year} {2016})}\BibitemShut {NoStop}%
\bibitem [{\citenamefont {Braunstein}\ and\ \citenamefont
  {Caves}(1994)}]{braunsteinStatisticalDistanceGeometry1994}%
  \BibitemOpen
  \bibfield  {author} {\bibinfo {author} {\bibfnamefont {S.~L.}\ \bibnamefont
  {Braunstein}}\ and\ \bibinfo {author} {\bibfnamefont {C.~M.}\ \bibnamefont
  {Caves}},\ }\bibfield  {title} {\bibinfo {title} {Statistical distance and
  the geometry of quantum states},\ }\href
  {https://doi.org/10.1103/PhysRevLett.72.3439} {\bibfield  {journal} {\bibinfo
   {journal} {Physical Review Letters}\ }\textbf {\bibinfo {volume} {72}},\
  \bibinfo {pages} {3439} (\bibinfo {year} {1994})}\BibitemShut {NoStop}%
\bibitem [{\citenamefont {Braunstein}\ \emph {et~al.}(1996)\citenamefont
  {Braunstein}, \citenamefont {Caves},\ and\ \citenamefont
  {Milburn}}]{Braunstein1996}%
  \BibitemOpen
  \bibfield  {author} {\bibinfo {author} {\bibfnamefont {S.~L.}\ \bibnamefont
  {Braunstein}}, \bibinfo {author} {\bibfnamefont {C.~M.}\ \bibnamefont
  {Caves}},\ and\ \bibinfo {author} {\bibfnamefont {G.}~\bibnamefont
  {Milburn}},\ }\bibfield  {title} {\bibinfo {title} {{Generalized Uncertainty
  Relations: Theory, Examples, and Lorentz Invariance}},\ }\href
  {https://doi.org/https://doi.org/10.1006/aphy.1996.0040} {\bibfield
  {journal} {\bibinfo  {journal} {Annals of Physics}\ }\textbf {\bibinfo
  {volume} {247}},\ \bibinfo {pages} {135} (\bibinfo {year}
  {1996})}\BibitemShut {NoStop}%
\bibitem [{\citenamefont {Giovannetti}\ \emph {et~al.}(2011)\citenamefont
  {Giovannetti}, \citenamefont {Lloyd},\ and\ \citenamefont
  {Maccone}}]{Vittorio2011}%
  \BibitemOpen
  \bibfield  {author} {\bibinfo {author} {\bibfnamefont {V.}~\bibnamefont
  {Giovannetti}}, \bibinfo {author} {\bibfnamefont {S.}~\bibnamefont {Lloyd}},\
  and\ \bibinfo {author} {\bibfnamefont {L.}~\bibnamefont {Maccone}},\
  }\bibfield  {title} {\bibinfo {title} {{Advances in quantum metrology}},\
  }\href {https://doi.org/10.1038/nphoton.2011.35} {\bibfield  {journal}
  {\bibinfo  {journal} {Nature Photonics}\ }\textbf {\bibinfo {volume} {5}},\
  \bibinfo {pages} {222} (\bibinfo {year} {2011})}\BibitemShut {NoStop}%
\bibitem [{\citenamefont {Tóth}\ and\ \citenamefont
  {Apellaniz}(2014)}]{Toth_2014}%
  \BibitemOpen
  \bibfield  {author} {\bibinfo {author} {\bibfnamefont {G.}~\bibnamefont
  {Tóth}}\ and\ \bibinfo {author} {\bibfnamefont {I.}~\bibnamefont
  {Apellaniz}},\ }\bibfield  {title} {\bibinfo {title} {Quantum metrology from
  a quantum information science perspective},\ }\href
  {https://doi.org/10.1088/1751-8113/47/42/424006} {\bibfield  {journal}
  {\bibinfo  {journal} {Journal of Physics A: Mathematical and Theoretical}\
  }\textbf {\bibinfo {volume} {47}},\ \bibinfo {pages} {424006} (\bibinfo
  {year} {2014})}\BibitemShut {NoStop}%
\bibitem [{\citenamefont {Pirandola}\ \emph {et~al.}(2018)\citenamefont
  {Pirandola}, \citenamefont {Bardhan}, \citenamefont {Gehring}, \citenamefont
  {Weedbrook},\ and\ \citenamefont {Lloyd}}]{Pirandola}%
  \BibitemOpen
  \bibfield  {author} {\bibinfo {author} {\bibfnamefont {S.}~\bibnamefont
  {Pirandola}}, \bibinfo {author} {\bibfnamefont {B.~R.}\ \bibnamefont
  {Bardhan}}, \bibinfo {author} {\bibfnamefont {T.}~\bibnamefont {Gehring}},
  \bibinfo {author} {\bibfnamefont {C.}~\bibnamefont {Weedbrook}},\ and\
  \bibinfo {author} {\bibfnamefont {S.}~\bibnamefont {Lloyd}},\ }\bibfield
  {title} {\bibinfo {title} {Advances in photonic quantum sensing},\ }\href
  {https://doi.org/10.1038/s41566-018-0301-6} {\bibfield  {journal} {\bibinfo
  {journal} {Nature Photonics}\ }\textbf {\bibinfo {volume} {12}},\ \bibinfo
  {pages} {724} (\bibinfo {year} {2018})}\BibitemShut {NoStop}%
\bibitem [{\citenamefont {Strobel}\ \emph {et~al.}(2014)\citenamefont
  {Strobel}, \citenamefont {Muessel}, \citenamefont {Linnemann}, \citenamefont
  {Zibold}, \citenamefont {Hume}, \citenamefont {Pezzè}, \citenamefont
  {Smerzi},\ and\ \citenamefont {Oberthaler}}]{Strobel}%
  \BibitemOpen
  \bibfield  {author} {\bibinfo {author} {\bibfnamefont {H.}~\bibnamefont
  {Strobel}}, \bibinfo {author} {\bibfnamefont {W.}~\bibnamefont {Muessel}},
  \bibinfo {author} {\bibfnamefont {D.}~\bibnamefont {Linnemann}}, \bibinfo
  {author} {\bibfnamefont {T.}~\bibnamefont {Zibold}}, \bibinfo {author}
  {\bibfnamefont {D.~B.}\ \bibnamefont {Hume}}, \bibinfo {author}
  {\bibfnamefont {L.}~\bibnamefont {Pezzè}}, \bibinfo {author} {\bibfnamefont
  {A.}~\bibnamefont {Smerzi}},\ and\ \bibinfo {author} {\bibfnamefont {M.~K.}\
  \bibnamefont {Oberthaler}},\ }\bibfield  {title} {\bibinfo {title} {{Fisher
  information and entanglement of non-Gaussian spin states}},\ }\href
  {https://doi.org/10.1126/science.1250147} {\bibfield  {journal} {\bibinfo
  {journal} {Science}\ }\textbf {\bibinfo {volume} {345}},\ \bibinfo {pages}
  {424} (\bibinfo {year} {2014})}\BibitemShut {NoStop}%
\bibitem [{\citenamefont {Gross}\ \emph {et~al.}(2010)\citenamefont {Gross},
  \citenamefont {Zibold}, \citenamefont {Nicklas}, \citenamefont {Est{\`e}ve},\
  and\ \citenamefont {Oberthaler}}]{Gross2010}%
  \BibitemOpen
  \bibfield  {author} {\bibinfo {author} {\bibfnamefont {C.}~\bibnamefont
  {Gross}}, \bibinfo {author} {\bibfnamefont {T.}~\bibnamefont {Zibold}},
  \bibinfo {author} {\bibfnamefont {E.}~\bibnamefont {Nicklas}}, \bibinfo
  {author} {\bibfnamefont {J.}~\bibnamefont {Est{\`e}ve}},\ and\ \bibinfo
  {author} {\bibfnamefont {M.~K.}\ \bibnamefont {Oberthaler}},\ }\bibfield
  {title} {\bibinfo {title} {Nonlinear atom interferometer surpasses classical
  precision limit},\ }\href {https://doi.org/10.1038/nature08919} {\bibfield
  {journal} {\bibinfo  {journal} {Nature}\ }\textbf {\bibinfo {volume} {464}},\
  \bibinfo {pages} {1165} (\bibinfo {year} {2010})}\BibitemShut {NoStop}%
\bibitem [{\citenamefont {Pezz\`e}\ \emph {et~al.}(2018)\citenamefont
  {Pezz\`e}, \citenamefont {Smerzi}, \citenamefont {Oberthaler}, \citenamefont
  {Schmied},\ and\ \citenamefont {Treutlein}}]{RevModPhys.90.035005}%
  \BibitemOpen
  \bibfield  {author} {\bibinfo {author} {\bibfnamefont {L.}~\bibnamefont
  {Pezz\`e}}, \bibinfo {author} {\bibfnamefont {A.}~\bibnamefont {Smerzi}},
  \bibinfo {author} {\bibfnamefont {M.~K.}\ \bibnamefont {Oberthaler}},
  \bibinfo {author} {\bibfnamefont {R.}~\bibnamefont {Schmied}},\ and\ \bibinfo
  {author} {\bibfnamefont {P.}~\bibnamefont {Treutlein}},\ }\bibfield  {title}
  {\bibinfo {title} {Quantum metrology with nonclassical states of atomic
  ensembles},\ }\href {https://doi.org/10.1103/RevModPhys.90.035005} {\bibfield
   {journal} {\bibinfo  {journal} {Rev. Mod. Phys.}\ }\textbf {\bibinfo
  {volume} {90}},\ \bibinfo {pages} {035005} (\bibinfo {year}
  {2018})}\BibitemShut {NoStop}%
\bibitem [{\citenamefont {Cassens}\ \emph {et~al.}(2025)\citenamefont
  {Cassens}, \citenamefont {Meyer-Hoppe}, \citenamefont {Rasel},\ and\
  \citenamefont {Klempt}}]{Klempt}%
  \BibitemOpen
  \bibfield  {author} {\bibinfo {author} {\bibfnamefont {C.}~\bibnamefont
  {Cassens}}, \bibinfo {author} {\bibfnamefont {B.}~\bibnamefont
  {Meyer-Hoppe}}, \bibinfo {author} {\bibfnamefont {E.}~\bibnamefont {Rasel}},\
  and\ \bibinfo {author} {\bibfnamefont {C.}~\bibnamefont {Klempt}},\
  }\bibfield  {title} {\bibinfo {title} {{Entanglement-Enhanced Atomic
  Gravimeter}},\ }\href {https://doi.org/10.1103/PhysRevX.15.011029} {\bibfield
   {journal} {\bibinfo  {journal} {Phys. Rev. X}\ }\textbf {\bibinfo {volume}
  {15}},\ \bibinfo {pages} {011029} (\bibinfo {year} {2025})}\BibitemShut
  {NoStop}%
\bibitem [{\citenamefont {Li}\ \emph {et~al.}(2026)\citenamefont {Li},
  \citenamefont {Joosten}, \citenamefont {Baamara}, \citenamefont {Colciaghi},
  \citenamefont {Sinatra}, \citenamefont {Treutlein},\ and\ \citenamefont
  {Zibold}}]{Treutlein}%
  \BibitemOpen
  \bibfield  {author} {\bibinfo {author} {\bibfnamefont {Y.}~\bibnamefont
  {Li}}, \bibinfo {author} {\bibfnamefont {L.}~\bibnamefont {Joosten}},
  \bibinfo {author} {\bibfnamefont {Y.}~\bibnamefont {Baamara}}, \bibinfo
  {author} {\bibfnamefont {P.}~\bibnamefont {Colciaghi}}, \bibinfo {author}
  {\bibfnamefont {A.}~\bibnamefont {Sinatra}}, \bibinfo {author} {\bibfnamefont
  {P.}~\bibnamefont {Treutlein}},\ and\ \bibinfo {author} {\bibfnamefont
  {T.}~\bibnamefont {Zibold}},\ }\bibfield  {title} {\bibinfo {title}
  {Multiparameter estimation with an array of entangled atomic sensors},\
  }\href {https://doi.org/10.1126/science.adt2442} {\bibfield  {journal}
  {\bibinfo  {journal} {Science}\ }\textbf {\bibinfo {volume} {391}},\ \bibinfo
  {pages} {374} (\bibinfo {year} {2026})}\BibitemShut {NoStop}%
\bibitem [{\citenamefont {Szigeti}\ \emph {et~al.}(2021)\citenamefont
  {Szigeti}, \citenamefont {Hosten},\ and\ \citenamefont {Haine}}]{Szigeti}%
  \BibitemOpen
  \bibfield  {author} {\bibinfo {author} {\bibfnamefont {S.~S.}\ \bibnamefont
  {Szigeti}}, \bibinfo {author} {\bibfnamefont {O.}~\bibnamefont {Hosten}},\
  and\ \bibinfo {author} {\bibfnamefont {S.~A.}\ \bibnamefont {Haine}},\
  }\bibfield  {title} {\bibinfo {title} {{Improving cold-atom sensors with
  quantum entanglement: Prospects and challenges}},\ }\href
  {https://doi.org/10.1063/5.0050235} {\bibfield  {journal} {\bibinfo
  {journal} {Applied Physics Letters}\ }\textbf {\bibinfo {volume} {118}},\
  \bibinfo {pages} {140501} (\bibinfo {year} {2021})}\BibitemShut {NoStop}%
\bibitem [{\citenamefont {Sinatra}\ \emph {et~al.}(2009)\citenamefont
  {Sinatra}, \citenamefont {Castin},\ and\ \citenamefont
  {Witkowska}}]{SinatraCastin}%
  \BibitemOpen
  \bibfield  {author} {\bibinfo {author} {\bibfnamefont {A.}~\bibnamefont
  {Sinatra}}, \bibinfo {author} {\bibfnamefont {Y.}~\bibnamefont {Castin}},\
  and\ \bibinfo {author} {\bibfnamefont {E.}~\bibnamefont {Witkowska}},\
  }\bibfield  {title} {\bibinfo {title} {{Coherence time of a Bose-Einstein
  condensate}},\ }\href {https://doi.org/10.1103/PhysRevA.80.033614} {\bibfield
   {journal} {\bibinfo  {journal} {Phys. Rev. A}\ }\textbf {\bibinfo {volume}
  {80}},\ \bibinfo {pages} {033614} (\bibinfo {year} {2009})}\BibitemShut
  {NoStop}%
\bibitem [{\citenamefont {Grond}\ \emph {et~al.}(2010)\citenamefont {Grond},
  \citenamefont {Hohenester}, \citenamefont {Mazets},\ and\ \citenamefont
  {Schmiedmayer}}]{Grond}%
  \BibitemOpen
  \bibfield  {author} {\bibinfo {author} {\bibfnamefont {J.}~\bibnamefont
  {Grond}}, \bibinfo {author} {\bibfnamefont {U.}~\bibnamefont {Hohenester}},
  \bibinfo {author} {\bibfnamefont {I.}~\bibnamefont {Mazets}},\ and\ \bibinfo
  {author} {\bibfnamefont {J.}~\bibnamefont {Schmiedmayer}},\ }\bibfield
  {title} {\bibinfo {title} {{Atom interferometry with trapped Bose–Einstein
  condensates: impact of atom–atom interactions}},\ }\href
  {https://doi.org/10.1088/1367-2630/12/6/065036} {\bibfield  {journal}
  {\bibinfo  {journal} {New Journal of Physics}\ }\textbf {\bibinfo {volume}
  {12}},\ \bibinfo {pages} {065036} (\bibinfo {year} {2010})}\BibitemShut
  {NoStop}%
\bibitem [{\citenamefont {Zanardi}\ \emph {et~al.}(2008)\citenamefont
  {Zanardi}, \citenamefont {Paris},\ and\ \citenamefont
  {Campos~Venuti}}]{Zanardi}%
  \BibitemOpen
  \bibfield  {author} {\bibinfo {author} {\bibfnamefont {P.}~\bibnamefont
  {Zanardi}}, \bibinfo {author} {\bibfnamefont {M.~G.~A.}\ \bibnamefont
  {Paris}},\ and\ \bibinfo {author} {\bibfnamefont {L.}~\bibnamefont
  {Campos~Venuti}},\ }\bibfield  {title} {\bibinfo {title} {Quantum criticality
  as a resource for quantum estimation},\ }\href
  {https://doi.org/10.1103/PhysRevA.78.042105} {\bibfield  {journal} {\bibinfo
  {journal} {Phys. Rev. A}\ }\textbf {\bibinfo {volume} {78}},\ \bibinfo
  {pages} {042105} (\bibinfo {year} {2008})}\BibitemShut {NoStop}%
\bibitem [{\citenamefont {Fr\'erot}\ and\ \citenamefont
  {Roscilde}(2018)}]{frerot}%
  \BibitemOpen
  \bibfield  {author} {\bibinfo {author} {\bibfnamefont {I.}~\bibnamefont
  {Fr\'erot}}\ and\ \bibinfo {author} {\bibfnamefont {T.}~\bibnamefont
  {Roscilde}},\ }\bibfield  {title} {\bibinfo {title} {{Quantum Critical
  Metrology}},\ }\href {https://doi.org/10.1103/PhysRevLett.121.020402}
  {\bibfield  {journal} {\bibinfo  {journal} {Phys. Rev. Lett.}\ }\textbf
  {\bibinfo {volume} {121}},\ \bibinfo {pages} {020402} (\bibinfo {year}
  {2018})}\BibitemShut {NoStop}%
\bibitem [{\citenamefont {Montenegro}\ \emph {et~al.}(2025)\citenamefont
  {Montenegro}, \citenamefont {Mukhopadhyay}, \citenamefont {Yousefjani},
  \citenamefont {Sarkar}, \citenamefont {Mishra}, \citenamefont {Paris},\ and\
  \citenamefont {Bayat}}]{montenegro}%
  \BibitemOpen
  \bibfield  {author} {\bibinfo {author} {\bibfnamefont {V.}~\bibnamefont
  {Montenegro}}, \bibinfo {author} {\bibfnamefont {C.}~\bibnamefont
  {Mukhopadhyay}}, \bibinfo {author} {\bibfnamefont {R.}~\bibnamefont
  {Yousefjani}}, \bibinfo {author} {\bibfnamefont {S.}~\bibnamefont {Sarkar}},
  \bibinfo {author} {\bibfnamefont {U.}~\bibnamefont {Mishra}}, \bibinfo
  {author} {\bibfnamefont {M.~G.~A.}\ \bibnamefont {Paris}},\ and\ \bibinfo
  {author} {\bibfnamefont {A.}~\bibnamefont {Bayat}},\ }\bibfield  {title}
  {\bibinfo {title} {Review: Quantum metrology and sensing with many-body
  systems},\ }\href
  {https://doi.org/https://doi.org/10.1016/j.physrep.2025.05.005} {\bibfield
  {journal} {\bibinfo  {journal} {Physics Reports}\ }\textbf {\bibinfo {volume}
  {1134}},\ \bibinfo {pages} {1} (\bibinfo {year} {2025})}\BibitemShut
  {NoStop}%
\bibitem [{\citenamefont {Mihailescu}\ \emph {et~al.}(2026)\citenamefont
  {Mihailescu}, \citenamefont {Alushi}, \citenamefont {Di~Candia},
  \citenamefont {Felicetti},\ and\ \citenamefont
  {Gietka}}]{mihailescuCriticalQuantumSensing2026}%
  \BibitemOpen
  \bibfield  {author} {\bibinfo {author} {\bibfnamefont {G.}~\bibnamefont
  {Mihailescu}}, \bibinfo {author} {\bibfnamefont {U.}~\bibnamefont {Alushi}},
  \bibinfo {author} {\bibfnamefont {R.}~\bibnamefont {Di~Candia}}, \bibinfo
  {author} {\bibfnamefont {S.}~\bibnamefont {Felicetti}},\ and\ \bibinfo
  {author} {\bibfnamefont {K.}~\bibnamefont {Gietka}},\ }\bibfield  {title}
  {\bibinfo {title} {{Critical Quantum Sensing: A Tutorial on Parameter
  Estimation Near Quantum Phase Transitions}},\ }\href
  {https://doi.org/10.1103/v7mf-yh8n} {\bibfield  {journal} {\bibinfo
  {journal} {PRX Quantum}\ }\textbf {\bibinfo {volume} {7}},\ \bibinfo {pages}
  {020201} (\bibinfo {year} {2026})}\BibitemShut {NoStop}%
\bibitem [{\citenamefont {Ragole}\ and\ \citenamefont {Taylor}(2016)}]{Taylor}%
  \BibitemOpen
  \bibfield  {author} {\bibinfo {author} {\bibfnamefont {S.}~\bibnamefont
  {Ragole}}\ and\ \bibinfo {author} {\bibfnamefont {J.~M.}\ \bibnamefont
  {Taylor}},\ }\bibfield  {title} {\bibinfo {title} {{Interacting Atomic
  Interferometry for Rotation Sensing Approaching the Heisenberg Limit}},\
  }\href {https://doi.org/10.1103/PhysRevLett.117.203002} {\bibfield  {journal}
  {\bibinfo  {journal} {Phys. Rev. Lett.}\ }\textbf {\bibinfo {volume} {117}},\
  \bibinfo {pages} {203002} (\bibinfo {year} {2016})}\BibitemShut {NoStop}%
\bibitem [{\citenamefont {Luo}\ \emph {et~al.}(2017)\citenamefont {Luo},
  \citenamefont {Huang}, \citenamefont {Zhang},\ and\ \citenamefont
  {Lee}}]{luoHeisenberglimitedSagnacInterferometer2017}%
  \BibitemOpen
  \bibfield  {author} {\bibinfo {author} {\bibfnamefont {C.}~\bibnamefont
  {Luo}}, \bibinfo {author} {\bibfnamefont {J.}~\bibnamefont {Huang}}, \bibinfo
  {author} {\bibfnamefont {X.}~\bibnamefont {Zhang}},\ and\ \bibinfo {author}
  {\bibfnamefont {C.}~\bibnamefont {Lee}},\ }\bibfield  {title} {\bibinfo
  {title} {Heisenberg-limited {{Sagnac}} interferometer with multiparticle
  states},\ }\href {https://doi.org/10.1103/PhysRevA.95.023608} {\bibfield
  {journal} {\bibinfo  {journal} {Physical Review A}\ }\textbf {\bibinfo
  {volume} {95}},\ \bibinfo {pages} {023608} (\bibinfo {year}
  {2017})}\BibitemShut {NoStop}%
\bibitem [{\citenamefont {Ou}(1997)}]{Ou}%
  \BibitemOpen
  \bibfield  {author} {\bibinfo {author} {\bibfnamefont {Z.~Y.}\ \bibnamefont
  {Ou}},\ }\bibfield  {title} {\bibinfo {title} {Fundamental quantum limit in
  precision phase measurement},\ }\href
  {https://doi.org/10.1103/PhysRevA.55.2598} {\bibfield  {journal} {\bibinfo
  {journal} {Phys. Rev. A}\ }\textbf {\bibinfo {volume} {55}},\ \bibinfo
  {pages} {2598} (\bibinfo {year} {1997})}\BibitemShut {NoStop}%
\bibitem [{\citenamefont {Giovannetti}\ \emph {et~al.}(2006)\citenamefont
  {Giovannetti}, \citenamefont {Lloyd},\ and\ \citenamefont
  {Maccone}}]{QuantumMetrology}%
  \BibitemOpen
  \bibfield  {author} {\bibinfo {author} {\bibfnamefont {V.}~\bibnamefont
  {Giovannetti}}, \bibinfo {author} {\bibfnamefont {S.}~\bibnamefont {Lloyd}},\
  and\ \bibinfo {author} {\bibfnamefont {L.}~\bibnamefont {Maccone}},\
  }\bibfield  {title} {\bibinfo {title} {{Quantum Metrology}},\ }\href
  {https://doi.org/10.1103/PhysRevLett.96.010401} {\bibfield  {journal}
  {\bibinfo  {journal} {Phys. Rev. Lett.}\ }\textbf {\bibinfo {volume} {96}},\
  \bibinfo {pages} {010401} (\bibinfo {year} {2006})}\BibitemShut {NoStop}%
\bibitem [{\citenamefont {Demkowicz-Dobrza{\'n}ski}\ \emph
  {et~al.}(2012)\citenamefont {Demkowicz-Dobrza{\'n}ski}, \citenamefont
  {Ko{\l}ody{\'n}ski},\ and\ \citenamefont {Gu{\c t}{\u a}}}]{Demkowicz}%
  \BibitemOpen
  \bibfield  {author} {\bibinfo {author} {\bibfnamefont {R.}~\bibnamefont
  {Demkowicz-Dobrza{\'n}ski}}, \bibinfo {author} {\bibfnamefont
  {J.}~\bibnamefont {Ko{\l}ody{\'n}ski}},\ and\ \bibinfo {author}
  {\bibfnamefont {M.}~\bibnamefont {Gu{\c t}{\u a}}},\ }\bibfield  {title}
  {\bibinfo {title} {{The elusive Heisenberg limit in quantum-enhanced
  metrology}},\ }\href {https://doi.org/10.1038/ncomms2067} {\bibfield
  {journal} {\bibinfo  {journal} {Nature Communications}\ }\textbf {\bibinfo
  {volume} {3}},\ \bibinfo {pages} {1063} (\bibinfo {year} {2012})}\BibitemShut
  {NoStop}%
\bibitem [{\citenamefont {Boixo}\ \emph {et~al.}(2009)\citenamefont {Boixo},
  \citenamefont {Datta}, \citenamefont {Davis}, \citenamefont {Shaji},
  \citenamefont {Tacla},\ and\ \citenamefont {Caves}}]{Boixo}%
  \BibitemOpen
  \bibfield  {author} {\bibinfo {author} {\bibfnamefont {S.}~\bibnamefont
  {Boixo}}, \bibinfo {author} {\bibfnamefont {A.}~\bibnamefont {Datta}},
  \bibinfo {author} {\bibfnamefont {M.~J.}\ \bibnamefont {Davis}}, \bibinfo
  {author} {\bibfnamefont {A.}~\bibnamefont {Shaji}}, \bibinfo {author}
  {\bibfnamefont {A.~B.}\ \bibnamefont {Tacla}},\ and\ \bibinfo {author}
  {\bibfnamefont {C.~M.}\ \bibnamefont {Caves}},\ }\bibfield  {title} {\bibinfo
  {title} {{Quantum-limited metrology and Bose-Einstein condensates}},\ }\href
  {https://doi.org/10.1103/PhysRevA.80.032103} {\bibfield  {journal} {\bibinfo
  {journal} {Phys. Rev. A}\ }\textbf {\bibinfo {volume} {80}},\ \bibinfo
  {pages} {032103} (\bibinfo {year} {2009})}\BibitemShut {NoStop}%
\bibitem [{\citenamefont {Braun}\ \emph {et~al.}(2018)\citenamefont {Braun},
  \citenamefont {Adesso}, \citenamefont {Benatti}, \citenamefont {Floreanini},
  \citenamefont {Marzolino}, \citenamefont {Mitchell},\ and\ \citenamefont
  {Pirandola}}]{BraunRMP}%
  \BibitemOpen
  \bibfield  {author} {\bibinfo {author} {\bibfnamefont {D.}~\bibnamefont
  {Braun}}, \bibinfo {author} {\bibfnamefont {G.}~\bibnamefont {Adesso}},
  \bibinfo {author} {\bibfnamefont {F.}~\bibnamefont {Benatti}}, \bibinfo
  {author} {\bibfnamefont {R.}~\bibnamefont {Floreanini}}, \bibinfo {author}
  {\bibfnamefont {U.}~\bibnamefont {Marzolino}}, \bibinfo {author}
  {\bibfnamefont {M.~W.}\ \bibnamefont {Mitchell}},\ and\ \bibinfo {author}
  {\bibfnamefont {S.}~\bibnamefont {Pirandola}},\ }\bibfield  {title} {\bibinfo
  {title} {Quantum-enhanced measurements without entanglement},\ }\href
  {https://doi.org/10.1103/RevModPhys.90.035006} {\bibfield  {journal}
  {\bibinfo  {journal} {Rev. Mod. Phys.}\ }\textbf {\bibinfo {volume} {90}},\
  \bibinfo {pages} {035006} (\bibinfo {year} {2018})}\BibitemShut {NoStop}%
\bibitem [{\citenamefont {Tsatsos}\ and\ \citenamefont
  {Lode}(2015)}]{tsatsosResonancesDynamicalFragmentation2015}%
  \BibitemOpen
  \bibfield  {author} {\bibinfo {author} {\bibfnamefont {M.~C.}\ \bibnamefont
  {Tsatsos}}\ and\ \bibinfo {author} {\bibfnamefont {A.~U.~J.}\ \bibnamefont
  {Lode}},\ }\bibfield  {title} {\bibinfo {title} {Resonances and {{Dynamical
  Fragmentation}} in a {{Stirred Bose}}--{{Einstein Condensate}}},\ }\href
  {https://doi.org/10.1007/s10909-015-1335-5} {\bibfield  {journal} {\bibinfo
  {journal} {Journal of Low Temperature Physics}\ }\textbf {\bibinfo {volume}
  {181}},\ \bibinfo {pages} {171} (\bibinfo {year} {2015})}\BibitemShut
  {NoStop}%
\bibitem [{\citenamefont {Beinke}\ \emph {et~al.}(2015)\citenamefont {Beinke},
  \citenamefont {Klaiman}, \citenamefont {Cederbaum}, \citenamefont
  {Streltsov},\ and\ \citenamefont
  {Alon}}]{beinkeManybodyTunnelingDynamics2015}%
  \BibitemOpen
  \bibfield  {author} {\bibinfo {author} {\bibfnamefont {R.}~\bibnamefont
  {Beinke}}, \bibinfo {author} {\bibfnamefont {S.}~\bibnamefont {Klaiman}},
  \bibinfo {author} {\bibfnamefont {L.~S.}\ \bibnamefont {Cederbaum}}, \bibinfo
  {author} {\bibfnamefont {A.~I.}\ \bibnamefont {Streltsov}},\ and\ \bibinfo
  {author} {\bibfnamefont {O.~E.}\ \bibnamefont {Alon}},\ }\bibfield  {title}
  {\bibinfo {title} {Many-body tunneling dynamics of {{Bose-Einstein}}
  condensates and vortex states in two spatial dimensions},\ }\href
  {https://doi.org/10.1103/PhysRevA.92.043627} {\bibfield  {journal} {\bibinfo
  {journal} {Physical Review A}\ }\textbf {\bibinfo {volume} {92}},\ \bibinfo
  {pages} {043627} (\bibinfo {year} {2015})}\BibitemShut {NoStop}%
\bibitem [{\citenamefont {Doganov}\ \emph {et~al.}(2013)\citenamefont
  {Doganov}, \citenamefont {Klaiman}, \citenamefont {Alon}, \citenamefont
  {Streltsov},\ and\ \citenamefont
  {Cederbaum}}]{doganovTwoTrappedParticles2013a}%
  \BibitemOpen
  \bibfield  {author} {\bibinfo {author} {\bibfnamefont {R.~A.}\ \bibnamefont
  {Doganov}}, \bibinfo {author} {\bibfnamefont {S.}~\bibnamefont {Klaiman}},
  \bibinfo {author} {\bibfnamefont {O.~E.}\ \bibnamefont {Alon}}, \bibinfo
  {author} {\bibfnamefont {A.~I.}\ \bibnamefont {Streltsov}},\ and\ \bibinfo
  {author} {\bibfnamefont {L.~S.}\ \bibnamefont {Cederbaum}},\ }\bibfield
  {title} {\bibinfo {title} {Two trapped particles interacting by a
  finite-range two-body potential in two spatial dimensions},\ }\href
  {https://doi.org/10.1103/PhysRevA.87.033631} {\bibfield  {journal} {\bibinfo
  {journal} {Physical Review A}\ }\textbf {\bibinfo {volume} {87}},\ \bibinfo
  {pages} {033631} (\bibinfo {year} {2013})}\BibitemShut {NoStop}%
\bibitem [{\citenamefont {Bhowmik}\ \emph {et~al.}(2020)\citenamefont
  {Bhowmik}, \citenamefont {Haldar},\ and\ \citenamefont
  {Alon}}]{bhowmikImpactTransverseDirection2020}%
  \BibitemOpen
  \bibfield  {author} {\bibinfo {author} {\bibfnamefont {A.}~\bibnamefont
  {Bhowmik}}, \bibinfo {author} {\bibfnamefont {S.~K.}\ \bibnamefont
  {Haldar}},\ and\ \bibinfo {author} {\bibfnamefont {O.~E.}\ \bibnamefont
  {Alon}},\ }\bibfield  {title} {\bibinfo {title} {Impact of the transverse
  direction on the many-body tunneling dynamics in a two-dimensional bosonic
  {{Josephson}} junction},\ }\href {https://doi.org/10.1038/s41598-020-78173-w}
  {\bibfield  {journal} {\bibinfo  {journal} {Scientific Reports}\ }\textbf
  {\bibinfo {volume} {10}},\ \bibinfo {pages} {21476} (\bibinfo {year}
  {2020})}\BibitemShut {NoStop}%
\bibitem [{\citenamefont {Pethick}\ and\ \citenamefont
  {Smith}(2008)}]{PethickSmith2008}%
  \BibitemOpen
  \bibfield  {author} {\bibinfo {author} {\bibfnamefont {C.~J.}\ \bibnamefont
  {Pethick}}\ and\ \bibinfo {author} {\bibfnamefont {H.}~\bibnamefont
  {Smith}},\ }\href {https://doi.org/10.1017/CBO9780511802850} {\emph {\bibinfo
  {title} {{Bose--Einstein Condensation in Dilute Gases}}}},\ \bibinfo
  {edition} {2nd}\ ed.\ (\bibinfo  {publisher} {Cambridge University Press},\
  \bibinfo {address} {Cambridge},\ \bibinfo {year} {2008})\BibitemShut
  {NoStop}%
\bibitem [{\citenamefont {Paredes}\ \emph {et~al.}(2004)\citenamefont
  {Paredes}, \citenamefont {Widera}, \citenamefont {Murg}, \citenamefont
  {Mandel}, \citenamefont {F{\"o}lling}, \citenamefont {Cirac}, \citenamefont
  {Shlyapnikov}, \citenamefont {H{\"a}nsch},\ and\ \citenamefont
  {Bloch}}]{Paredes}%
  \BibitemOpen
  \bibfield  {author} {\bibinfo {author} {\bibfnamefont {B.}~\bibnamefont
  {Paredes}}, \bibinfo {author} {\bibfnamefont {A.}~\bibnamefont {Widera}},
  \bibinfo {author} {\bibfnamefont {V.}~\bibnamefont {Murg}}, \bibinfo {author}
  {\bibfnamefont {O.}~\bibnamefont {Mandel}}, \bibinfo {author} {\bibfnamefont
  {S.}~\bibnamefont {F{\"o}lling}}, \bibinfo {author} {\bibfnamefont
  {I.}~\bibnamefont {Cirac}}, \bibinfo {author} {\bibfnamefont {G.~V.}\
  \bibnamefont {Shlyapnikov}}, \bibinfo {author} {\bibfnamefont {T.~W.}\
  \bibnamefont {H{\"a}nsch}},\ and\ \bibinfo {author} {\bibfnamefont
  {I.}~\bibnamefont {Bloch}},\ }\bibfield  {title} {\bibinfo {title}
  {{Tonks--Girardeau gas of ultracold atoms in an optical lattice}},\ }\href
  {https://doi.org/10.1038/nature02530} {\bibfield  {journal} {\bibinfo
  {journal} {Nature}\ }\textbf {\bibinfo {volume} {429}},\ \bibinfo {pages}
  {277} (\bibinfo {year} {2004})}\BibitemShut {NoStop}%
\bibitem [{\citenamefont {Dengis}\ \emph {et~al.}(2026)\citenamefont {Dengis},
  \citenamefont {Dupont}, \citenamefont {Schlagheck},\ and\ \citenamefont
  {Goldman}}]{dengisVortexNOONStates2026}%
  \BibitemOpen
  \bibfield  {author} {\bibinfo {author} {\bibfnamefont {S.}~\bibnamefont
  {Dengis}}, \bibinfo {author} {\bibfnamefont {N.}~\bibnamefont {Dupont}},
  \bibinfo {author} {\bibfnamefont {P.}~\bibnamefont {Schlagheck}},\ and\
  \bibinfo {author} {\bibfnamefont {N.}~\bibnamefont {Goldman}},\ }\href
  {https://doi.org/10.48550/arXiv.2606.29509} {\bibinfo {title} {Vortex
  {{NOON}} states for rotation sensing}} (\bibinfo {year} {2026}),\ \Eprint
  {https://arxiv.org/abs/2606.29509} {2606.29509 [cond-mat.quant-gas]}
  \BibitemShut {NoStop}%
\bibitem [{\citenamefont
  {Helstrom}(1967)}]{helstromMinimumMeansquaredError1967b}%
  \BibitemOpen
  \bibfield  {author} {\bibinfo {author} {\bibfnamefont {C.}~\bibnamefont
  {Helstrom}},\ }\bibfield  {title} {\bibinfo {title} {Minimum mean-squared
  error of estimates in quantum statistics},\ }\href
  {https://doi.org/10.1016/0375-9601(67)90366-0} {\bibfield  {journal}
  {\bibinfo  {journal} {Physics Letters A}\ }\textbf {\bibinfo {volume} {25}},\
  \bibinfo {pages} {101} (\bibinfo {year} {1967})}\BibitemShut {NoStop}%
\bibitem [{\citenamefont {Facchi}\ \emph {et~al.}(2010)\citenamefont {Facchi},
  \citenamefont {Kulkarni}, \citenamefont {Man'ko}, \citenamefont {Marmo},
  \citenamefont {Sudarshan},\ and\ \citenamefont
  {Ventriglia}}]{facchiClassicalQuantumFisher2010}%
  \BibitemOpen
  \bibfield  {author} {\bibinfo {author} {\bibfnamefont {P.}~\bibnamefont
  {Facchi}}, \bibinfo {author} {\bibfnamefont {R.}~\bibnamefont {Kulkarni}},
  \bibinfo {author} {\bibfnamefont {V.}~\bibnamefont {Man'ko}}, \bibinfo
  {author} {\bibfnamefont {G.}~\bibnamefont {Marmo}}, \bibinfo {author}
  {\bibfnamefont {E.}~\bibnamefont {Sudarshan}},\ and\ \bibinfo {author}
  {\bibfnamefont {F.}~\bibnamefont {Ventriglia}},\ }\bibfield  {title}
  {\bibinfo {title} {Classical and quantum {{Fisher}} information in the
  geometrical formulation of quantum mechanics},\ }\href
  {https://doi.org/10.1016/j.physleta.2010.10.005} {\bibfield  {journal}
  {\bibinfo  {journal} {Physics Letters A}\ }\textbf {\bibinfo {volume}
  {374}},\ \bibinfo {pages} {4801} (\bibinfo {year} {2010})}\BibitemShut
  {NoStop}%
\bibitem [{\citenamefont {Milburn}\ \emph {et~al.}(1997)\citenamefont
  {Milburn}, \citenamefont {Corney}, \citenamefont {Wright},\ and\
  \citenamefont {Walls}}]{milburnQuantumDynamicsAtomic1997a}%
  \BibitemOpen
  \bibfield  {author} {\bibinfo {author} {\bibfnamefont {G.~J.}\ \bibnamefont
  {Milburn}}, \bibinfo {author} {\bibfnamefont {J.}~\bibnamefont {Corney}},
  \bibinfo {author} {\bibfnamefont {E.~M.}\ \bibnamefont {Wright}},\ and\
  \bibinfo {author} {\bibfnamefont {D.~F.}\ \bibnamefont {Walls}},\ }\bibfield
  {title} {\bibinfo {title} {Quantum dynamics of an atomic {{Bose-Einstein}}
  condensate in a double-well potential},\ }\href
  {https://doi.org/10.1103/PhysRevA.55.4318} {\bibfield  {journal} {\bibinfo
  {journal} {Physical Review A}\ }\textbf {\bibinfo {volume} {55}},\ \bibinfo
  {pages} {4318} (\bibinfo {year} {1997})}\BibitemShut {NoStop}%
\bibitem [{\citenamefont {Dalton}\ and\ \citenamefont
  {Ghanbari}(2012)}]{daltonTwoModeTheory2012}%
  \BibitemOpen
  \bibfield  {author} {\bibinfo {author} {\bibfnamefont {B.}~\bibnamefont
  {Dalton}}\ and\ \bibinfo {author} {\bibfnamefont {S.}~\bibnamefont
  {Ghanbari}},\ }\bibfield  {title} {\bibinfo {title} {Two mode theory of
  {{Bose}}--{{Einstein}} condensates: Interferometry and the {{Josephson}}
  model},\ }\href {https://doi.org/10.1080/09500340.2011.632100} {\bibfield
  {journal} {\bibinfo  {journal} {Journal of Modern Optics}\ }\textbf {\bibinfo
  {volume} {59}},\ \bibinfo {pages} {287} (\bibinfo {year} {2012})}\BibitemShut
  {NoStop}%
\bibitem [{\citenamefont {Naldesi}\ \emph {et~al.}(2022)\citenamefont
  {Naldesi}, \citenamefont {Polo}, \citenamefont {Dunjko}, \citenamefont
  {Perrin}, \citenamefont {Olshanii}, \citenamefont {Amico},\ and\
  \citenamefont {Minguzzi}}]{naldesiEnhancingSensitivityRotations2022}%
  \BibitemOpen
  \bibfield  {author} {\bibinfo {author} {\bibfnamefont {P.}~\bibnamefont
  {Naldesi}}, \bibinfo {author} {\bibfnamefont {J.}~\bibnamefont {Polo}},
  \bibinfo {author} {\bibfnamefont {V.}~\bibnamefont {Dunjko}}, \bibinfo
  {author} {\bibfnamefont {H.}~\bibnamefont {Perrin}}, \bibinfo {author}
  {\bibfnamefont {M.}~\bibnamefont {Olshanii}}, \bibinfo {author}
  {\bibfnamefont {L.}~\bibnamefont {Amico}},\ and\ \bibinfo {author}
  {\bibfnamefont {A.}~\bibnamefont {Minguzzi}},\ }\bibfield  {title} {\bibinfo
  {title} {{Enhancing Sensitivity to Rotations with Quantum Solitonic
  Currents}},\ }\href {https://doi.org/10.21468/SciPostPhys.12.4.138}
  {\bibfield  {journal} {\bibinfo  {journal} {SciPost Physics}\ }\textbf
  {\bibinfo {volume} {12}},\ \bibinfo {pages} {138} (\bibinfo {year}
  {2022})}\BibitemShut {NoStop}%
\bibitem [{\citenamefont {Polo}\ \emph
  {et~al.}(2025{\natexlab{b}})\citenamefont {Polo}, \citenamefont {Chetcuti},
  \citenamefont {Haug}, \citenamefont {Minguzzi}, \citenamefont {Wright},\ and\
  \citenamefont {Amico}}]{poloPersistentCurrentsUltracold2025}%
  \BibitemOpen
  \bibfield  {author} {\bibinfo {author} {\bibfnamefont {J.}~\bibnamefont
  {Polo}}, \bibinfo {author} {\bibfnamefont {W.}~\bibnamefont {Chetcuti}},
  \bibinfo {author} {\bibfnamefont {T.}~\bibnamefont {Haug}}, \bibinfo {author}
  {\bibfnamefont {A.}~\bibnamefont {Minguzzi}}, \bibinfo {author}
  {\bibfnamefont {K.}~\bibnamefont {Wright}},\ and\ \bibinfo {author}
  {\bibfnamefont {L.}~\bibnamefont {Amico}},\ }\bibfield  {title} {\bibinfo
  {title} {{Persistent Currents in Ultracold Gases}},\ }\href
  {https://doi.org/10.1016/j.physrep.2025.06.003} {\bibfield  {journal}
  {\bibinfo  {journal} {Physics Reports}\ }\textbf {\bibinfo {volume} {1137}},\
  \bibinfo {pages} {1} (\bibinfo {year} {2025}{\natexlab{b}})}\BibitemShut
  {NoStop}%
\bibitem [{\citenamefont {Jaksch}\ \emph {et~al.}(1998)\citenamefont {Jaksch},
  \citenamefont {Bruder}, \citenamefont {Cirac}, \citenamefont {Gardiner},\
  and\ \citenamefont {Zoller}}]{jakschColdBosonicAtoms1998a}%
  \BibitemOpen
  \bibfield  {author} {\bibinfo {author} {\bibfnamefont {D.}~\bibnamefont
  {Jaksch}}, \bibinfo {author} {\bibfnamefont {C.}~\bibnamefont {Bruder}},
  \bibinfo {author} {\bibfnamefont {J.~I.}\ \bibnamefont {Cirac}}, \bibinfo
  {author} {\bibfnamefont {C.~W.}\ \bibnamefont {Gardiner}},\ and\ \bibinfo
  {author} {\bibfnamefont {P.}~\bibnamefont {Zoller}},\ }\bibfield  {title}
  {\bibinfo {title} {Cold {{Bosonic Atoms}} in {{Optical Lattices}}},\ }\href
  {https://doi.org/10.1103/PhysRevLett.81.3108} {\bibfield  {journal} {\bibinfo
   {journal} {Physical Review Letters}\ }\textbf {\bibinfo {volume} {81}},\
  \bibinfo {pages} {3108} (\bibinfo {year} {1998})}\BibitemShut {NoStop}%
\bibitem [{\citenamefont {Greiner}\ \emph {et~al.}(2002)\citenamefont
  {Greiner}, \citenamefont {Mandel}, \citenamefont {Esslinger}, \citenamefont
  {H{\"a}nsch},\ and\ \citenamefont {Bloch}}]{Greiner}%
  \BibitemOpen
  \bibfield  {author} {\bibinfo {author} {\bibfnamefont {M.}~\bibnamefont
  {Greiner}}, \bibinfo {author} {\bibfnamefont {O.}~\bibnamefont {Mandel}},
  \bibinfo {author} {\bibfnamefont {T.}~\bibnamefont {Esslinger}}, \bibinfo
  {author} {\bibfnamefont {T.~W.}\ \bibnamefont {H{\"a}nsch}},\ and\ \bibinfo
  {author} {\bibfnamefont {I.}~\bibnamefont {Bloch}},\ }\bibfield  {title}
  {\bibinfo {title} {{Quantum phase transition from a superfluid to a Mott
  insulator in a gas of ultracold atoms}},\ }\href
  {https://doi.org/10.1038/415039a} {\bibfield  {journal} {\bibinfo  {journal}
  {Nature}\ }\textbf {\bibinfo {volume} {415}},\ \bibinfo {pages} {39}
  (\bibinfo {year} {2002})}\BibitemShut {NoStop}%
\bibitem [{\citenamefont {Sachdev}(2011)}]{sachdevQuantumPhaseTransitions2011}%
  \BibitemOpen
  \bibfield  {author} {\bibinfo {author} {\bibfnamefont {S.}~\bibnamefont
  {Sachdev}},\ }\href@noop {} {\emph {\bibinfo {title} {Quantum Phase
  Transitions}}},\ \bibinfo {edition} {2nd}\ ed.\ (\bibinfo  {publisher}
  {Cambridge University Press},\ \bibinfo {year} {2011})\BibitemShut {NoStop}%
\bibitem [{\citenamefont {Meyer}\ \emph {et~al.}(1990)\citenamefont {Meyer},
  \citenamefont {Manthe},\ and\ \citenamefont {Cederbaum}}]{Meyer}%
  \BibitemOpen
  \bibfield  {author} {\bibinfo {author} {\bibfnamefont {H.-D.}\ \bibnamefont
  {Meyer}}, \bibinfo {author} {\bibfnamefont {U.}~\bibnamefont {Manthe}},\ and\
  \bibinfo {author} {\bibfnamefont {L.}~\bibnamefont {Cederbaum}},\ }\bibfield
  {title} {\bibinfo {title} {{The multi-configurational time-dependent Hartree
  approach}},\ }\href
  {https://doi.org/https://doi.org/10.1016/0009-2614(90)87014-I} {\bibfield
  {journal} {\bibinfo  {journal} {Chemical Physics Letters}\ }\textbf {\bibinfo
  {volume} {165}},\ \bibinfo {pages} {73} (\bibinfo {year} {1990})}\BibitemShut
  {NoStop}%
\bibitem [{\citenamefont {Streltsov}\ \emph {et~al.}(2006)\citenamefont
  {Streltsov}, \citenamefont {Alon},\ and\ \citenamefont
  {Cederbaum}}]{streltsovGeneralVariationalManybody2006}%
  \BibitemOpen
  \bibfield  {author} {\bibinfo {author} {\bibfnamefont {A.~I.}\ \bibnamefont
  {Streltsov}}, \bibinfo {author} {\bibfnamefont {O.~E.}\ \bibnamefont
  {Alon}},\ and\ \bibinfo {author} {\bibfnamefont {L.~S.}\ \bibnamefont
  {Cederbaum}},\ }\bibfield  {title} {\bibinfo {title} {{General Variational
  Many-Body Theory with Complete Self-Consistency for Trapped Bosonic
  Systems}},\ }\href {https://doi.org/10.1103/PhysRevA.73.063626} {\bibfield
  {journal} {\bibinfo  {journal} {Physical Review A}\ }\textbf {\bibinfo
  {volume} {73}},\ \bibinfo {pages} {063626} (\bibinfo {year}
  {2006})}\BibitemShut {NoStop}%
\bibitem [{\citenamefont {Alon}\ \emph {et~al.}(2008)\citenamefont {Alon},
  \citenamefont {Streltsov},\ and\ \citenamefont
  {Cederbaum}}]{alonMulticonfigurationalTimedependentHartree2008}%
  \BibitemOpen
  \bibfield  {author} {\bibinfo {author} {\bibfnamefont {O.~E.}\ \bibnamefont
  {Alon}}, \bibinfo {author} {\bibfnamefont {A.~I.}\ \bibnamefont
  {Streltsov}},\ and\ \bibinfo {author} {\bibfnamefont {L.~S.}\ \bibnamefont
  {Cederbaum}},\ }\bibfield  {title} {\bibinfo {title} {Multiconfigurational
  time-dependent {{Hartree}} method for bosons: {{Many-body}} dynamics of
  bosonic systems},\ }\href {https://doi.org/10.1103/PhysRevA.77.033613}
  {\bibfield  {journal} {\bibinfo  {journal} {Physical Review A}\ }\textbf
  {\bibinfo {volume} {77}},\ \bibinfo {pages} {033613} (\bibinfo {year}
  {2008})}\BibitemShut {NoStop}%
\bibitem [{\citenamefont {Lode}\ \emph {et~al.}(2020)\citenamefont {Lode},
  \citenamefont {L{\'e}v{\^e}que}, \citenamefont {Madsen}, \citenamefont
  {Streltsov},\ and\ \citenamefont
  {Alon}}]{lodeColloquiumMulticonfigurationalTimedependent2020}%
  \BibitemOpen
  \bibfield  {author} {\bibinfo {author} {\bibfnamefont {A.~U.~J.}\
  \bibnamefont {Lode}}, \bibinfo {author} {\bibfnamefont {C.}~\bibnamefont
  {L{\'e}v{\^e}que}}, \bibinfo {author} {\bibfnamefont {L.~B.}\ \bibnamefont
  {Madsen}}, \bibinfo {author} {\bibfnamefont {A.~I.}\ \bibnamefont
  {Streltsov}},\ and\ \bibinfo {author} {\bibfnamefont {O.~E.}\ \bibnamefont
  {Alon}},\ }\bibfield  {title} {\bibinfo {title} {Colloquium:
  {{Multiconfigurational}} time-dependent {{Hartree}} approaches for
  indistinguishable particles},\ }\href
  {https://doi.org/10.1103/RevModPhys.92.011001} {\bibfield  {journal}
  {\bibinfo  {journal} {Reviews of Modern Physics}\ }\textbf {\bibinfo {volume}
  {92}},\ \bibinfo {pages} {011001} (\bibinfo {year} {2020})}\BibitemShut
  {NoStop}%
\bibitem [{\citenamefont {Roy}\ and\ \citenamefont {Alon}(2025)}]{Roy}%
  \BibitemOpen
  \bibfield  {author} {\bibinfo {author} {\bibfnamefont {R.}~\bibnamefont
  {Roy}}\ and\ \bibinfo {author} {\bibfnamefont {O.~E.}\ \bibnamefont {Alon}},\
  }\bibfield  {title} {\bibinfo {title} {{Assessing small accelerations using a
  bosonic Josephson junction}},\ }\href
  {https://doi.org/10.1103/PhysRevA.111.043307} {\bibfield  {journal} {\bibinfo
   {journal} {Phys. Rev. A}\ }\textbf {\bibinfo {volume} {111}},\ \bibinfo
  {pages} {043307} (\bibinfo {year} {2025})}\BibitemShut {NoStop}%
\bibitem [{\citenamefont {Roy}\ and\ \citenamefont
  {Alon}(2026)}]{Roy_rotation}%
  \BibitemOpen
  \bibfield  {author} {\bibinfo {author} {\bibfnamefont {R.}~\bibnamefont
  {Roy}}\ and\ \bibinfo {author} {\bibfnamefont {O.~E.}\ \bibnamefont {Alon}},\
  }\href {https://arxiv.org/abs/2601.13344} {\bibinfo {title} {{Inferring
  rotations using a bosonic Josephson junction}}} (\bibinfo {year} {2026}),\
  \Eprint {https://arxiv.org/abs/2601.13344} {arXiv:2601.13344
  [cond-mat.quant-gas]} \BibitemShut {NoStop}%
\bibitem [{\citenamefont {Lin}\ \emph {et~al.}(2020)\citenamefont {Lin},
  \citenamefont {Molignini}, \citenamefont {Papariello}, \citenamefont
  {Tsatsos}, \citenamefont {L{\'e}v{\^e}que}, \citenamefont {Weiner},
  \citenamefont {Fasshauer}, \citenamefont {Chitra},\ and\ \citenamefont
  {Lode}}]{linMCTDHXMulticonfigurationalTimedependent2020}%
  \BibitemOpen
  \bibfield  {author} {\bibinfo {author} {\bibfnamefont {R.}~\bibnamefont
  {Lin}}, \bibinfo {author} {\bibfnamefont {P.}~\bibnamefont {Molignini}},
  \bibinfo {author} {\bibfnamefont {L.}~\bibnamefont {Papariello}}, \bibinfo
  {author} {\bibfnamefont {M.~C.}\ \bibnamefont {Tsatsos}}, \bibinfo {author}
  {\bibfnamefont {C.}~\bibnamefont {L{\'e}v{\^e}que}}, \bibinfo {author}
  {\bibfnamefont {S.~E.}\ \bibnamefont {Weiner}}, \bibinfo {author}
  {\bibfnamefont {E.}~\bibnamefont {Fasshauer}}, \bibinfo {author}
  {\bibfnamefont {R.}~\bibnamefont {Chitra}},\ and\ \bibinfo {author}
  {\bibfnamefont {A.~U.~J.}\ \bibnamefont {Lode}},\ }\bibfield  {title}
  {\bibinfo {title} {{{MCTDH-X}}: {{The}} multiconfigurational time-dependent
  {{Hartree}} method for indistinguishable particles software},\ }\href
  {https://doi.org/10.1088/2058-9565/ab788b} {\bibfield  {journal} {\bibinfo
  {journal} {{Quantum Science and Technology}}\ }\textbf {\bibinfo {volume}
  {5}},\ \bibinfo {pages} {024004} (\bibinfo {year} {2020})}\BibitemShut
  {NoStop}%
\bibitem [{\citenamefont {Gwak}\ \emph {et~al.}(2021)\citenamefont {Gwak},
  \citenamefont {Marchukov},\ and\ \citenamefont {Fischer}}]{Gwak}%
  \BibitemOpen
  \bibfield  {author} {\bibinfo {author} {\bibfnamefont {Y.}~\bibnamefont
  {Gwak}}, \bibinfo {author} {\bibfnamefont {O.~V.}\ \bibnamefont
  {Marchukov}},\ and\ \bibinfo {author} {\bibfnamefont {U.~R.}\ \bibnamefont
  {Fischer}},\ }\bibfield  {title} {\bibinfo {title} {{Benchmarking the
  multiconfigurational Hartree method by the exact wavefunction of two
  harmonically trapped bosons with contact interaction}},\ }\href
  {https://doi.org/https://doi.org/10.1016/j.aop.2021.168592} {\bibfield
  {journal} {\bibinfo  {journal} {Annals of Physics}\ }\textbf {\bibinfo
  {volume} {434}},\ \bibinfo {pages} {168592} (\bibinfo {year}
  {2021})}\BibitemShut {NoStop}%
\bibitem [{\citenamefont {Baak}\ and\ \citenamefont
  {Fischer}(2024)}]{baakSelfConsistentManyBodyMetrology2024}%
  \BibitemOpen
  \bibfield  {author} {\bibinfo {author} {\bibfnamefont {J.-G.}\ \bibnamefont
  {Baak}}\ and\ \bibinfo {author} {\bibfnamefont {U.~R.}\ \bibnamefont
  {Fischer}},\ }\bibfield  {title} {\bibinfo {title} {Self-{{Consistent
  Many-Body Metrology}}},\ }\href
  {https://doi.org/10.1103/PhysRevLett.132.240803} {\bibfield  {journal}
  {\bibinfo  {journal} {Physical Review Letters}\ }\textbf {\bibinfo {volume}
  {132}},\ \bibinfo {pages} {240803} (\bibinfo {year} {2024})}\BibitemShut
  {NoStop}%
\bibitem [{\citenamefont {Sakmann}\ \emph {et~al.}(2009)\citenamefont
  {Sakmann}, \citenamefont {Streltsov}, \citenamefont {Alon},\ and\
  \citenamefont {Cederbaum}}]{Sakmann}%
  \BibitemOpen
  \bibfield  {author} {\bibinfo {author} {\bibfnamefont {K.}~\bibnamefont
  {Sakmann}}, \bibinfo {author} {\bibfnamefont {A.~I.}\ \bibnamefont
  {Streltsov}}, \bibinfo {author} {\bibfnamefont {O.~E.}\ \bibnamefont
  {Alon}},\ and\ \bibinfo {author} {\bibfnamefont {L.~S.}\ \bibnamefont
  {Cederbaum}},\ }\bibfield  {title} {\bibinfo {title} {{Exact Quantum Dynamics
  of a Bosonic Josephson Junction}},\ }\href
  {https://doi.org/10.1103/PhysRevLett.103.220601} {\bibfield  {journal}
  {\bibinfo  {journal} {Phys. Rev. Lett.}\ }\textbf {\bibinfo {volume} {103}},\
  \bibinfo {pages} {220601} (\bibinfo {year} {2009})}\BibitemShut {NoStop}%
\bibitem [{\citenamefont {Klaiman}\ and\ \citenamefont {Alon}(2015)}]{Klaiman}%
  \BibitemOpen
  \bibfield  {author} {\bibinfo {author} {\bibfnamefont {S.}~\bibnamefont
  {Klaiman}}\ and\ \bibinfo {author} {\bibfnamefont {O.~E.}\ \bibnamefont
  {Alon}},\ }\bibfield  {title} {\bibinfo {title} {Variance as a sensitive
  probe of correlations},\ }\href {https://doi.org/10.1103/PhysRevA.91.063613}
  {\bibfield  {journal} {\bibinfo  {journal} {Phys. Rev. A}\ }\textbf {\bibinfo
  {volume} {91}},\ \bibinfo {pages} {063613} (\bibinfo {year}
  {2015})}\BibitemShut {NoStop}%
\bibitem [{\citenamefont {Glauber}(1963)}]{glauberQuantumTheoryOptical1963}%
  \BibitemOpen
  \bibfield  {author} {\bibinfo {author} {\bibfnamefont {R.~J.}\ \bibnamefont
  {Glauber}},\ }\bibfield  {title} {\bibinfo {title} {The {{Quantum Theory}} of
  {{Optical Coherence}}},\ }\href {https://doi.org/10.1103/PhysRev.130.2529}
  {\bibfield  {journal} {\bibinfo  {journal} {Physical Review}\ }\textbf
  {\bibinfo {volume} {130}},\ \bibinfo {pages} {2529} (\bibinfo {year}
  {1963})}\BibitemShut {NoStop}%
\bibitem [{\citenamefont {Roy}\ \emph {et~al.}(2018)\citenamefont {Roy},
  \citenamefont {Gammal}, \citenamefont {Tsatsos}, \citenamefont {Chatterjee},
  \citenamefont {Chakrabarti},\ and\ \citenamefont
  {Lode}}]{royPhasesManybodyEntropy2018}%
  \BibitemOpen
  \bibfield  {author} {\bibinfo {author} {\bibfnamefont {R.}~\bibnamefont
  {Roy}}, \bibinfo {author} {\bibfnamefont {A.}~\bibnamefont {Gammal}},
  \bibinfo {author} {\bibfnamefont {M.~C.}\ \bibnamefont {Tsatsos}}, \bibinfo
  {author} {\bibfnamefont {B.}~\bibnamefont {Chatterjee}}, \bibinfo {author}
  {\bibfnamefont {B.}~\bibnamefont {Chakrabarti}},\ and\ \bibinfo {author}
  {\bibfnamefont {A.~U.~J.}\ \bibnamefont {Lode}},\ }\bibfield  {title}
  {\bibinfo {title} {Phases, many-body entropy measures, and coherence of
  interacting bosons in optical lattices},\ }\href
  {https://doi.org/10.1103/PhysRevA.97.043625} {\bibfield  {journal} {\bibinfo
  {journal} {Physical Review A}\ }\textbf {\bibinfo {volume} {97}},\ \bibinfo
  {pages} {043625} (\bibinfo {year} {2018})}\BibitemShut {NoStop}%
\bibitem [{\citenamefont {Lin}\ \emph {et~al.}(2019)\citenamefont {Lin},
  \citenamefont {Papariello}, \citenamefont {Molignini}, \citenamefont
  {Chitra},\ and\ \citenamefont {Lode}}]{linSuperfluidMottInsulator2019}%
  \BibitemOpen
  \bibfield  {author} {\bibinfo {author} {\bibfnamefont {R.}~\bibnamefont
  {Lin}}, \bibinfo {author} {\bibfnamefont {L.}~\bibnamefont {Papariello}},
  \bibinfo {author} {\bibfnamefont {P.}~\bibnamefont {Molignini}}, \bibinfo
  {author} {\bibfnamefont {R.}~\bibnamefont {Chitra}},\ and\ \bibinfo {author}
  {\bibfnamefont {A.~U.~J.}\ \bibnamefont {Lode}},\ }\bibfield  {title}
  {\bibinfo {title} {Superfluid--{{Mott-insulator}} transition of ultracold
  superradiant bosons in a cavity},\ }\href
  {https://doi.org/10.1103/PhysRevA.100.013611} {\bibfield  {journal} {\bibinfo
   {journal} {Physical Review A}\ }\textbf {\bibinfo {volume} {100}},\ \bibinfo
  {pages} {013611} (\bibinfo {year} {2019})}\BibitemShut {NoStop}%
\bibitem [{\citenamefont {Bo{\'e}ris}\ \emph {et~al.}(2016)\citenamefont
  {Bo{\'e}ris}, \citenamefont {Gori}, \citenamefont {Hoogerland}, \citenamefont
  {Kumar}, \citenamefont {Lucioni}, \citenamefont {Tanzi}, \citenamefont
  {Inguscio}, \citenamefont {Giamarchi}, \citenamefont {D'Errico},
  \citenamefont {Carleo}, \citenamefont {Modugno},\ and\ \citenamefont
  {{Sanchez-Palencia}}}]{boerisMottTransitionStrongly2016}%
  \BibitemOpen
  \bibfield  {author} {\bibinfo {author} {\bibfnamefont {G.}~\bibnamefont
  {Bo{\'e}ris}}, \bibinfo {author} {\bibfnamefont {L.}~\bibnamefont {Gori}},
  \bibinfo {author} {\bibfnamefont {M.~D.}\ \bibnamefont {Hoogerland}},
  \bibinfo {author} {\bibfnamefont {A.}~\bibnamefont {Kumar}}, \bibinfo
  {author} {\bibfnamefont {E.}~\bibnamefont {Lucioni}}, \bibinfo {author}
  {\bibfnamefont {L.}~\bibnamefont {Tanzi}}, \bibinfo {author} {\bibfnamefont
  {M.}~\bibnamefont {Inguscio}}, \bibinfo {author} {\bibfnamefont
  {T.}~\bibnamefont {Giamarchi}}, \bibinfo {author} {\bibfnamefont
  {C.}~\bibnamefont {D'Errico}}, \bibinfo {author} {\bibfnamefont
  {G.}~\bibnamefont {Carleo}}, \bibinfo {author} {\bibfnamefont
  {G.}~\bibnamefont {Modugno}},\ and\ \bibinfo {author} {\bibfnamefont
  {L.}~\bibnamefont {{Sanchez-Palencia}}},\ }\bibfield  {title} {\bibinfo
  {title} {{Mott Transition for Strongly Interacting One-Dimensional Bosons in
  a Shallow Periodic Potential}},\ }\href
  {https://doi.org/10.1103/PhysRevA.93.011601} {\bibfield  {journal} {\bibinfo
  {journal} {Physical Review A}\ }\textbf {\bibinfo {volume} {93}},\ \bibinfo
  {pages} {011601} (\bibinfo {year} {2016})}\BibitemShut {NoStop}%
\bibitem [{\citenamefont {Alon}\ and\ \citenamefont
  {Cederbaum}(2005)}]{alonPathwayCondensationFragmentation2005}%
  \BibitemOpen
  \bibfield  {author} {\bibinfo {author} {\bibfnamefont {O.~E.}\ \bibnamefont
  {Alon}}\ and\ \bibinfo {author} {\bibfnamefont {L.~S.}\ \bibnamefont
  {Cederbaum}},\ }\bibfield  {title} {\bibinfo {title} {Pathway from
  {{Condensation}} via {{Fragmentation}} to {{Fermionization}} of {{Cold
  Bosonic Systems}}},\ }\href {https://doi.org/10.1103/PhysRevLett.95.140402}
  {\bibfield  {journal} {\bibinfo  {journal} {Physical Review Letters}\
  }\textbf {\bibinfo {volume} {95}},\ \bibinfo {pages} {140402} (\bibinfo
  {year} {2005})}\BibitemShut {NoStop}%
\bibitem [{\citenamefont {Baak}\ and\ \citenamefont
  {Fischer}(2022)}]{LiebLinigerMetro}%
  \BibitemOpen
  \bibfield  {author} {\bibinfo {author} {\bibfnamefont {J.-G.}\ \bibnamefont
  {Baak}}\ and\ \bibinfo {author} {\bibfnamefont {U.~R.}\ \bibnamefont
  {Fischer}},\ }\bibfield  {title} {\bibinfo {title} {{Classical and quantum
  metrology of the Lieb-Liniger model}},\ }\href
  {https://doi.org/10.1103/PhysRevA.106.062442} {\bibfield  {journal} {\bibinfo
   {journal} {Phys. Rev. A}\ }\textbf {\bibinfo {volume} {106}},\ \bibinfo
  {pages} {062442} (\bibinfo {year} {2022})}\BibitemShut {NoStop}%
\bibitem [{\citenamefont {Girardeau}(1960)}]{TG}%
  \BibitemOpen
  \bibfield  {author} {\bibinfo {author} {\bibfnamefont {M.}~\bibnamefont
  {Girardeau}},\ }\bibfield  {title} {\bibinfo {title} {{Relationship between
  Systems of Impenetrable Bosons and Fermions in One Dimension}},\ }\href
  {https://doi.org/10.1063/1.1703687} {\bibfield  {journal} {\bibinfo
  {journal} {Journal of Mathematical Physics}\ }\textbf {\bibinfo {volume}
  {1}},\ \bibinfo {pages} {516} (\bibinfo {year} {1960})}\BibitemShut {NoStop}%
\bibitem [{\citenamefont {Bakr}\ \emph {et~al.}(2009)\citenamefont {Bakr},
  \citenamefont {Gillen}, \citenamefont {Peng}, \citenamefont {F{\"o}lling},\
  and\ \citenamefont {Greiner}}]{Bakr}%
  \BibitemOpen
  \bibfield  {author} {\bibinfo {author} {\bibfnamefont {W.~S.}\ \bibnamefont
  {Bakr}}, \bibinfo {author} {\bibfnamefont {J.~I.}\ \bibnamefont {Gillen}},
  \bibinfo {author} {\bibfnamefont {A.}~\bibnamefont {Peng}}, \bibinfo {author}
  {\bibfnamefont {S.}~\bibnamefont {F{\"o}lling}},\ and\ \bibinfo {author}
  {\bibfnamefont {M.}~\bibnamefont {Greiner}},\ }\bibfield  {title} {\bibinfo
  {title} {{A quantum gas microscope for detecting single atoms in a
  Hubbard-regime optical lattice}},\ }\href
  {https://doi.org/10.1038/nature08482} {\bibfield  {journal} {\bibinfo
  {journal} {Nature}\ }\textbf {\bibinfo {volume} {462}},\ \bibinfo {pages}
  {74} (\bibinfo {year} {2009})}\BibitemShut {NoStop}%
\bibitem [{\citenamefont {Gross}\ and\ \citenamefont {Bakr}(2021)}]{Gross}%
  \BibitemOpen
  \bibfield  {author} {\bibinfo {author} {\bibfnamefont {C.}~\bibnamefont
  {Gross}}\ and\ \bibinfo {author} {\bibfnamefont {W.~S.}\ \bibnamefont
  {Bakr}},\ }\bibfield  {title} {\bibinfo {title} {Quantum gas microscopy for
  single atom and spin detection},\ }\href
  {https://doi.org/10.1038/s41567-021-01370-5} {\bibfield  {journal} {\bibinfo
  {journal} {Nature Physics}\ }\textbf {\bibinfo {volume} {17}},\ \bibinfo
  {pages} {1316} (\bibinfo {year} {2021})}\BibitemShut {NoStop}%
\bibitem [{\citenamefont {Yao}\ \emph {et~al.}(2025)\citenamefont {Yao},
  \citenamefont {Chi}, \citenamefont {Wang}, \citenamefont {Fletcher},\ and\
  \citenamefont {Zwierlein}}]{Zwierlein}%
  \BibitemOpen
  \bibfield  {author} {\bibinfo {author} {\bibfnamefont {R.}~\bibnamefont
  {Yao}}, \bibinfo {author} {\bibfnamefont {S.}~\bibnamefont {Chi}}, \bibinfo
  {author} {\bibfnamefont {M.}~\bibnamefont {Wang}}, \bibinfo {author}
  {\bibfnamefont {R.~J.}\ \bibnamefont {Fletcher}},\ and\ \bibinfo {author}
  {\bibfnamefont {M.}~\bibnamefont {Zwierlein}},\ }\bibfield  {title} {\bibinfo
  {title} {{Measuring Pair Correlations in Bose and Fermi Gases via
  Atom-Resolved Microscopy}},\ }\href
  {https://doi.org/10.1103/PhysRevLett.134.183402} {\bibfield  {journal}
  {\bibinfo  {journal} {Phys. Rev. Lett.}\ }\textbf {\bibinfo {volume} {134}},\
  \bibinfo {pages} {183402} (\bibinfo {year} {2025})}\BibitemShut {NoStop}%
\bibitem [{\citenamefont {Amico}\ \emph {et~al.}(2022)\citenamefont {Amico},
  \citenamefont {Anderson}, \citenamefont {Boshier}, \citenamefont {Brantut},
  \citenamefont {Kwek}, \citenamefont {Minguzzi},\ and\ \citenamefont {von
  Klitzing}}]{Amico}%
  \BibitemOpen
  \bibfield  {author} {\bibinfo {author} {\bibfnamefont {L.}~\bibnamefont
  {Amico}}, \bibinfo {author} {\bibfnamefont {D.}~\bibnamefont {Anderson}},
  \bibinfo {author} {\bibfnamefont {M.}~\bibnamefont {Boshier}}, \bibinfo
  {author} {\bibfnamefont {J.-P.}\ \bibnamefont {Brantut}}, \bibinfo {author}
  {\bibfnamefont {L.-C.}\ \bibnamefont {Kwek}}, \bibinfo {author}
  {\bibfnamefont {A.}~\bibnamefont {Minguzzi}},\ and\ \bibinfo {author}
  {\bibfnamefont {W.}~\bibnamefont {von Klitzing}},\ }\bibfield  {title}
  {\bibinfo {title} {{Colloquium: Atomtronic circuits: From many-body physics
  to quantum technologies}},\ }\href
  {https://doi.org/10.1103/RevModPhys.94.041001} {\bibfield  {journal}
  {\bibinfo  {journal} {Rev. Mod. Phys.}\ }\textbf {\bibinfo {volume} {94}},\
  \bibinfo {pages} {041001} (\bibinfo {year} {2022})}\BibitemShut {NoStop}%
\bibitem [{\citenamefont {H\"am\"al\"ainen}\ \emph {et~al.}(1993)\citenamefont
  {H\"am\"al\"ainen}, \citenamefont {Hari}, \citenamefont {Ilmoniemi},
  \citenamefont {Knuutila},\ and\ \citenamefont {Lounasmaa}}]{Lounasmaa}%
  \BibitemOpen
  \bibfield  {author} {\bibinfo {author} {\bibfnamefont {M.}~\bibnamefont
  {H\"am\"al\"ainen}}, \bibinfo {author} {\bibfnamefont {R.}~\bibnamefont
  {Hari}}, \bibinfo {author} {\bibfnamefont {R.~J.}\ \bibnamefont {Ilmoniemi}},
  \bibinfo {author} {\bibfnamefont {J.}~\bibnamefont {Knuutila}},\ and\
  \bibinfo {author} {\bibfnamefont {O.~V.}\ \bibnamefont {Lounasmaa}},\
  }\bibfield  {title} {\bibinfo {title} {Magnetoencephalography---theory,
  instrumentation, and applications to noninvasive studies of the working human
  brain},\ }\href {https://doi.org/10.1103/RevModPhys.65.413} {\bibfield
  {journal} {\bibinfo  {journal} {Rev. Mod. Phys.}\ }\textbf {\bibinfo {volume}
  {65}},\ \bibinfo {pages} {413} (\bibinfo {year} {1993})}\BibitemShut
  {NoStop}%
\bibitem [{Note1()}]{Note1}%
  \BibitemOpen
  \bibinfo {note} {We note that Ref.~\cite
  {baakSelfConsistentManyBodyMetrology2024} omitted a term $ \DOTSB \sum@
  \slimits@ _{k,q}\langle \chi _k|\chi _q\rangle \rho _{kq}=\DOTSB \sum@
  \slimits@ _{k,q}(\partial ^2_X)_{kq}\rho _{kq}+\DOTSB \sum@ \slimits@
  _{k,s,q}(\partial _X)_{ks}(\partial _X)_{sq}\rho _{kq}$ in the QFI, which
  corresponds to the change of orbitals orthogonal to the subspace of the $M$
  computational modes.}\BibitemShut {Stop}%
\bibitem [{\citenamefont {Penrose}\ and\ \citenamefont
  {Onsager}(1956)}]{PenroseOnsagerBEC1956}%
  \BibitemOpen
  \bibfield  {author} {\bibinfo {author} {\bibfnamefont {O.}~\bibnamefont
  {Penrose}}\ and\ \bibinfo {author} {\bibfnamefont {L.}~\bibnamefont
  {Onsager}},\ }\bibfield  {title} {\bibinfo {title} {{Bose-Einstein
  Condensation and Liquid Helium}},\ }\href
  {https://doi.org/10.1103/PhysRev.104.576} {\bibfield  {journal} {\bibinfo
  {journal} {Phys. Rev.}\ }\textbf {\bibinfo {volume} {104}},\ \bibinfo {pages}
  {576} (\bibinfo {year} {1956})}\BibitemShut {NoStop}%
\end{thebibliography}%





\begin{appendices}


\section*{End Matter}

\subsection{A: QFI in the MCTDH framework} 
The many-body wave function ansatz in the MCTDH framework for a system of interacting bosons is given as a linear combination of time-dependent permanents $\ket{\vec{n}; t}$, using the conventions of Ref.~\cite{alonMulticonfigurationalTimedependentHartree2008}
\begin{eqnarray}
    \ket{\Psi(t)}&=&\sum_{\vec{n}}C_{\vec{n}}(t)\ket{\vec{n};t},\quad \vec{n}=(n_1, \dots, n_M)^T, \nn
    \ket{\vec{n};t}&=&\frac{1}{\sqrt{\prod_{k=1}^{M}n_k!}}\prod_{k=1}^M\left[\hat{b}^\dagger_k(t)\right]^{n_k}\ket{0},
    \label{MCTDHAnsatz}
\end{eqnarray}
where $\{C_{\vec{n}}(t)\}$ are the expansion coefficients, $\hat{b}_k(t)=\int d\mathbf{r}\phi_k^*(\mathbf{r};t)\hat{\mathrm{\Psi}}(\mathbf{r})$ are the bosonic annihilation operators corresponding for the (orthonormalized) 
orbitals $\{\phi_k(\mathbf{r};t):k=1,\dots,M\}$, and the summation runs over all possible occupations that preserve the total particle number $n_1+\dots+n_M=N$. 
The reduced one-body density matrix elements $\rho_{kq}=\bra{\Psi}\hat{b}^\dagger_k\hat{b}_q\ket{\Psi}$ become
\begin{multline}
    \rho_{kk}=\sum_{\vec{n}}|C_{\vec{n}}|^2n_k,\quad
    \rho_{kq}=\sum_{\vec{n}}C^*_{\vec{n}}C_{\vec{n}_k^q}\sqrt{n_k(n_q+1)},
\end{multline}
where by convention different indices $k,q$ assume different values and $\vec{n}_k^q$ is the distribution obtained from $\vec{n}$ by (de-)exciting   
one boson from the $k$th to the $q$th orbital. 
We also define 
\begin{eqnarray}
    \zeta_{kq}=\begin{cases}
    n_k, & q = k,\\
    \sqrt{n_k(n_q+1)},  & q\ne k.
    \end{cases}
\end{eqnarray}
such that $\rho_{kq}=\sum_{\vec{n}}C^*_{\vec{n}}C_{\vec{n}_k^q}\zeta_{kq}$ for all $k,q$. The elements of the reduced two-body density matrix $ \rho_{ksql}=\bra{\Psi}\hat{b}^\dagger_k\hat{b}^\dagger_s\hat{b}_q\hat{b}_l\ket{\Psi}$ can be similarly obtained as functionals of the coefficients and occupation 
numbers~\cite{alonMulticonfigurationalTimedependentHartree2008,baakSelfConsistentManyBodyMetrology2024}.

Considering a Hamiltonian parameter estimation problem, we assume a state $\ket{\Psi_X}$ dependent on a parameter $X$, on which thus both the coefficients and the orbitals will generally depend. Importantly, the 
$X$-dependence of the orbital is
\begin{eqnarray}
    \ket{\partial_X\phi_q}=\sum_k(\partial_X)_{kq}\ket{\phi_k}+\ket{\chi_q}
\end{eqnarray}
where $(\partial_X)_{kq}=\int d\mathbf{r}\phi^*_k(\mathbf{r})\partial_X\phi_q(\mathbf{r})$, and $\ket{\chi_q}=\hat{\mathbf{P}}\ket{\partial_X\phi_q}$ is the component orthogonal to the subspace spanned by the $M$ orbitals, given by the projection operator $\hat{\mathbf{P}}=\mathbf{1}-\sum_k\ket{\phi_k}\bra{\phi_k}$. In second-quantized form, the derivative acting on the permanents can be written as
\begin{eqnarray}
\label{eq:dperm}
    \partial_X\ket{\vec{n}}=\sum_{kq}(\partial_X)_{kq}\hat{b}_k^\dagger\hat{b}_q\ket{\vec{n}}+\sum_{q}\hat{c}^\dagger_q\hat{b}_q\ket{\vec{n}}
\end{eqnarray}
where $\hat{c}_q=\int d\mathbf{r}\chi^*_q(\mathbf{r})\hat{\Psi}(\mathbf{r})$.  The QFI \eqref{FQpure}
has thus contributions from the parametric dependencies of coefficients, orbitals, and a cross term 
as follows 
\begin{eqnarray}
\label{eq:qfi_total}
    \mathcal{F}_Q=\mathcal{F}_Q^\mathrm{(coef)}+\mathcal{F}_Q^\mathrm{(cross)}+\mathcal{F}_Q^\mathrm{(orb)}.
\end{eqnarray}
Plugging in the MCTDH ansatz \eqref{MCTDHAnsatz}, we obtain 
\begin{eqnarray}
\label{eq:qfi_coef}
\frac14 \mathcal{F}_Q^\mathrm{(coef)}&=&\sum_{\vec{n}}\left|\partial_XC_{\vec{n}}\right|^2-\left|\sum_{\vec{n}}C_{\vec{n}}^*\partial_XC_{\vec{n}}\right|^2,
    \nn
\label{eq:qfi_cross}
 \frac14  \mathcal{F}_Q^\mathrm{(cross)}&=&\sum_{\vec{n}}\sum_{k,q}\left(C_{\vec{n}_k^q}\partial_XC_{\vec{n}}^*-C_{\vec{n}}^*\partial_XC_{\vec{n}_k^q}\right)(\partial_X)_{kq}\zeta_{kq} \nn
  &  &  -\sum_{\vec{n}}\left(C_{\vec{n}}\partial_XC_{\vec{n}}^*-C_{\vec{n}}^*\partial_XC_{\vec{n}}\right)\sum_{k,q}(\partial_X)_{kq}\rho_{kq}, \nn
\label{eq:qfi_orb}
  \frac14  \mathcal{F}_Q^\mathrm{(orb)} &=&\sum_{k,q}(\partial^2_X)_{kq}\rho_{kq}-\sum_{k,s,q,l}(\partial_X)_{kq}(\partial_X)_{sl}\rho_{ksql}\nn
&&   -\left|\sum_{k,q}(\partial_X)_{kq}\rho_{kq}\right|^2,
\end{eqnarray}
abbreviating $(\partial^2_X)_{kq}\coloneqq \int d\mathbf{r}\partial_X\phi^*_k(\mathbf{r})\partial_X\phi_q(\mathbf{r})$ 
\footnote{We note that Ref.~\cite{baakSelfConsistentManyBodyMetrology2024} omitted a term 
$ \sum_{k,q}\braket{\chi_k}{\chi_q}\rho_{kq}=\sum_{k,q}(\partial^2_X)_{kq}\rho_{kq}+\sum_{k,s,q}(\partial_X)_{ks}(\partial_X)_{sq}\rho_{kq}$ in the QFI, 
which corresponds to the change of orbitals orthogonal to the subspace of the $M$ computational modes. 
}. 

\subsection{B: Occupation entropy and ``fermionization"} 
We consider 
commensurate filling, i.e., take $N$ even. 
We introduce the occupation entropy (taking the form of a Shannon information measure)  
$S_\mathrm{occ}$, which is defined as \cite{royPhasesManybodyEntropy2018} 
\begin{eqnarray}
    S_\mathrm{occ}=-\sum_{i=1}^{M}n_i\ln{n_i}, \label{Socc}
\end{eqnarray}
where $n_i$ are the eigenvalues of the reduced one-body density matrix $\rho^{(1)}$ (normalized to unity, 
$\sum_{i=1}^M n_i =1$), which provide 
the canonical measure of fragmentation for interacting bosons according to Penrose and Onsager~\cite{PenroseOnsagerBEC1956}. 
A Bose-Einstein condensate, characterized by a single macroscopic 
eigenvalue of $\rho^{(1)}$ of order unity (formally defined in the thermodynamic limit  $N\rightarrow \infty$) 
 implies $S_\mathrm{occ}=0$,  
while $S_\mathrm{occ}$ increases and saturates at its maximal (ground state) value $\ln N$  as the system becomes 
fully fragmented. 
This corresponds to a  ``fermionization" in each well  for large couplings 
\cite{alonPathwayCondensationFragmentation2005}. 
\begin{figure}[b]
    \centering
     \includegraphics[width=\linewidth]{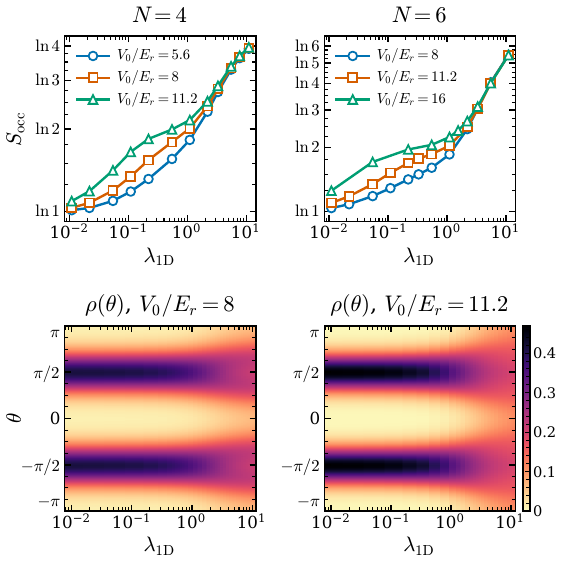}
    \caption{First row: Occupation entropy \eqref{Socc} as a function of coupling $\lambda_{\rm 1D}$.  
    The normalized one-body density $\rho(\theta)$ 
    along the azimuthal direction is displayed in the second row for $V_0/E_r=8$ ($N=4$) and $V_0/E_r=11.2$ ($N=6$).
    $M=12$ ($M=18$) orbitals were used for $N=4$ ($N=6$).} 
    \label{fig:qfi_scan_lambda_N_4_6}
\end{figure}

As the system enters the ``fermionized" (Tonks-Girardeau) limit, bosons initially localized in the same well tend to occupy different orbitals; the density $\rho(\theta)$ acquires gradual oscillations, see Fig.~\ref{fig:qfi_scan_lambda_N_4_6}, second row. 
Then, $S_\mathrm{occ}\rightarrow\ln4$ 
($S_\mathrm{occ}\rightarrow\ln6$) for $N=4$ ($N=6$) in the fully fragmented ``fermionized" state (first row of
Fig.~\ref{fig:qfi_scan_lambda_N_4_6}), accompanied by a steep decrease of the QFI, cf.~Fig.~\ref{fig:qfi_corr_N_2}(b) in the main text. Such a behavior cannot be observed for $N=2$ particles, where also stronger interactions always help maximizing the QFI, since only a single boson is present in each lattice site upon 
entering the localized regime. 
\end{appendices}


\end{document}